\documentclass[aps, pre, a4paper, floatfix,  twocolumn, longbibliography, nofootinbib]{revtex4-1}

\usepackage[utf8]{inputenc}
\usepackage{epsfig}
\usepackage[T1]{fontenc}
\usepackage[english]{babel}
\usepackage[table]{xcolor}
\usepackage{t1enc}
\usepackage{graphicx}
\usepackage{amssymb}
\usepackage{amsmath}
\usepackage{relsize}
\usepackage[percent]{overpic}
\usepackage{bm}
\usepackage{hyperref}

\usepackage[normalem]{ulem}

\newcommand{\avg}[1]{{\left<#1\right>}}

\newcommand{\dd}{\mathrm{d}}
\newcommand{\ee}{\mathrm{e}}

\usepackage{mathtools}
\def\multiset#1#2{\ensuremath{\left(\kern-.3em\left(\genfrac{}{}{0pt}{}{#1}{#2}\right)\kern-.3em\right)}}

\usepackage{amsmath}

\newcommand{\A}{\bm{A}}

\newcommand{\bb}{\bm{b}}
\newcommand{\x}{{\bm{x}}}

\newcommand{\e}{\mathrm{e}}

\usepackage{verbatim}
\usepackage{overpic}
\usepackage{booktabs}
\usepackage{placeins}

\usepackage{siunitx}
\usepackage{multirow}

\begin{document}

\title{Merge-split Markov chain Monte Carlo for community detection}

\author{Tiago P. Peixoto}
\email{peixotot@ceu.edu}
\affiliation{Department of Network and Data Science, Central European University, H-1051 Budapest, Hungary}
\affiliation{ISI Foundation, Via Chisola 5, 10126 Torino, Italy}
\affiliation{Department of Mathematical Sciences, University of Bath, Claverton Down, Bath BA2
  7AY, United Kingdom}
\begin{abstract}
  We present a Markov chain Monte Carlo scheme based on merges and
  splits of groups that is capable of efficiently sampling from the
  posterior distribution of network partitions, defined according to the
  stochastic block model (SBM). We demonstrate how schemes based on the
  move of single nodes between groups systematically fail at correctly
  sampling from the posterior distribution even on small networks, and
  how our merge-split approach behaves significantly better, and
  improves the mixing time of the Markov chain by several orders of
  magnitude in typical cases.  We also show how the scheme can be
  straightforwardly extended to nested versions of the SBM, yielding
  asymptotically exact samples of hierarchical network partitions.
\end{abstract}

\maketitle

\section{Introduction}

Community detection~\cite{fortunato_community_2010} is an important tool
for the analysis of network data, as it allows the breaking down of the
network structure into basic building blocks, thus providing a summary
of its large-scale structure. Among the many methods proposed for this
task, those based on the statistical inference of generative models
offer the most principled and robust
choice~\cite{peixoto_bayesian_2019}. This is due in large part to their
ability to take into account and convey the statistical evidence
available in the data, allowing us to prevent both overfitting ---
occurring when purely random fluctuations are mistaken by structure ---
as well as underfitting --- occurring when statistically significant
structure is mistaken by random fluctuations. These problems manifest
themselves in nonstatistical methods in their tendency to identify
spurious communities in fully random
networks~\cite{guimera_modularity_2004} as well as those that are
nonmodular but sparse~\cite{bagrow_communities_2012}, and in the
failure at recognizing the existence of clear but relatively small
communities in large networks~\cite{fortunato_resolution_2007}. The
manner in which statistical inference allows us to overcome these
obstacles is by allowing us to express not only our belief about how the
data are generated, but also about how much information is needed to
specify the model parameters. The combination of these two elements
defines a posterior distribution of all possible community structures
that are compatible with the data, ranked according to their
plausibility. The posterior distribution inherently ascribes low
probabilities to overly-complicated models, and yields the most
parsimonious but sufficiently rich explanation for the
data. Nevertheless, in case the model being used cannot perfectly
encapsulate the structure in the data, the posterior distribution may
still give comparatively high probabilities to multiple diverging
explanations for the data. Choosing only one fit of the model, for
example by selecting a single modular division with the strictly largest
posterior probability, is a form of bias in this situation. Instead, it
is preferable to consider the entire set of plausible partitions, by
sampling from the posterior distribution, rather than maximizing it.

Although it is relatively easy to write down a mathematical expression
for the posterior distribution of network partitions, at least up to a
normalization constant, using a simple model family called the
stochastic block model (SBM)~\cite{holland_stochastic_1983,
karrer_stochastic_2011}, actually sampling from this distribution is a
different matter altogether. In fact, fitting the SBM in this way
subsumes solving certain instances of combinatorial problems such as
graph coloring that are known to be NP-hard, and indeed there is strong
statistical evidence pointing to the existence of parameter regimes
where, although it may be possible to characterize the posterior
distribution of the SBM, this can only be done in nonpolynomial
time~\cite{decelle_asymptotic_2011}. Because of this inherent
intractability, it is unlikely that there will ever be an algorithm that
samples directly from the posterior distribution in an efficient manner,
at least not for networks that have more than a handful of nodes, as the
number of possible partitions grows super-exponentially with network
size. Nevertheless, we can still approach the problem in practice using
approximate methods. One attractive choice is Markov chain Monte Carlo
(MCMC), which consists in moving from one partition of the network to
another, with a probability conditioned only on the current state, in a
manner that guarantees exact asymptotic convergence to the desired
posterior distribution, provided one is able to wait long enough for the
chain to ``mix.'' In this way, one should be able to sample effectively from
the posterior in feasible instances of the problem, which would
correspond to a short mixing time.

In the case of the SBM, MCMC sampling has been implemented in a variety
of ways~\cite{snijders_estimation_1997,guimera_missing_2009,
  peixoto_efficient_2014,newman_estimating_2016,riolo_efficient_2017},
but all approaches thus far employed involve moves from one partition to
another that differ only in the group membership of a single node. As we
will show, these ``single-node'' approaches fail systematically at
efficiently exploring the posterior distribution of partitions, as they
cannot overcome low probability ``barriers'' that prevent the chain from
mixing. This is particularly relevant in situations where the number of
groups is unknown a priori, and the movement of a single node at a time
forces the Markov chain to move through low-probability states composed
of groups containing only one node, if the chain is to visit partitions
with different numbers of groups. These stopgap transitions prevent the
MCMC from mixing in a reasonable amount of time, and result in abysmal
performance even in easy instances of the problem.

In this work we present an alternative MCMC scheme that involves the
simultaneous movement of several nodes at each step, via the merging,
splitting and re-arrangements of groups. As we show, these moves can
easily overcome the low-probability barriers that block the single-move
approaches, and cause the chain to mix several orders of magnitude
faster. This improvement makes MCMC usable for significantly larger
networks, and allows us to explore community structure in empirical
networks with more confidence and increased amount of detail.

This paper is defined as follows. In Sec.~\ref{sec:bayes} we briefly
review the inference approach to community detection. In
Sec.~\ref{sec:mcmc} we describe the MCMC approach with single-node moves
and in Sec.~\ref{sec:mergesplit} we describe the version with merges and
splits. In Sec.~\ref{sec:empirical} we investigate the performance of
both algorithms in empirical networks, and in Sec.~\ref{sec:nested} we
extend the algorithm to hierarchical models. We end in
Sec.~\ref{sec:conc} with a conclusion.

\section{Bayesian inference of community structure}\label{sec:bayes}

The inference approach to community detection involves first the
stipulation of a generative model for the network structure given a
network partition $\bb=\{b_1,\dots,b_N\}$, where $b_i\in[1,B]$ is the
membership of node $i$ in one of $B$ groups. For example, when using the
degree-corrected stochastic block model
(DC-SBM)~\cite{karrer_stochastic_2011}, it is assumed that a network
with adjacency matrix $\A$ is generated with probability
\begin{multline}
  P(\A|\bm\lambda,\bm\theta,\bb) =
  \prod_{i<j}\frac{(\lambda_{b_ib_j}\theta_i\theta_j)^{A_{ij}}\e^{-\lambda_{b_ib_j}\theta_i\theta_j}}{A_{ij}!}\times\\ \prod_i\frac{(\lambda_{b_ib_i}\theta_i^2/2)^{A_{ij}/2}\e^{-\lambda_{b_ib_i}\theta_i^2/2}}{(A_{ij}/2)!},
\end{multline}
with the additional parameters $\bm\lambda$ and $\bm\theta$ specifying
the affinity between groups and expected node degrees, respectively. The
inference procedure consists in sampling from the posterior distribution
\begin{equation}\label{eq:posterior}
  P(\bb|\A) = \frac{P(\A|\bb)P(\bb)}{P(\A)},
\end{equation}
where $P(\bb)$ is a prior probability of node partitions, and
\begin{equation}
  P(\A|\bb) = \int P(\A|\bm\lambda,\bm\theta,\bb) P(\bm\theta|\bb) P(\bm\lambda|\bb)\;\dd\bm\theta\dd\bm\lambda,
\end{equation}
is the marginal likelihood integrated over the remaining model
parameters, weighted according to their own prior probabilities. If we
make a noninformative choice for them,
\begin{align}
  P(\bm\theta|\bb) &= \prod_r(n_r-1)!\,\delta\left(\textstyle\sum_i\theta_i\delta_{b_i,r}-1\right),\\
  P(\bm\lambda|\bb,\bar\lambda) &= \prod_{r<s}\e^{-\lambda_{rs}/\bar\lambda}/\bar\lambda\prod_r\e^{-\lambda_{rs}/2\bar\lambda}/2\bar\lambda,
\end{align}
we can compute the integral exactly as~\cite{peixoto_nonparametric_2017}
\begin{multline}\label{eq:marginal_likelihood}
  P(\A|\bb) = \frac{\bar\lambda^E}{(\bar\lambda+1)^{E+B(B+1)/2}}\times \\\frac{\prod_{r<s}e_{rs}!\prod_r e_{rr}!!}{\prod_{i<j}A_{ij}!\prod_i A_{ii}!!}
  \prod_r\frac{(n_r-1)!}{(e_r+n_r-1)!}\prod_ik_i!,
\end{multline}
where $e_{rs}=\sum_{ij}A_{ij}\delta_{b_i,r}\delta_{b_j,s}$,
$e_r=\sum_se_{rs}$, and $n_r=\sum_i\delta_{b_i,r}$. The last remaining
quantity
\begin{equation}
  P(\A) = \sum_{\bb}P(\A|\bb)P(\bb)
\end{equation}
is called the evidence, and servers as a normalization constant for the
posterior distribution. Although it has an important role in the context
of model selection, its value cannot be computed in closed form in the
general case. Luckily this is not needed for MCMC, as we will shortly
see.

The scheme above can be modified in a variety of ways, many of which can
significantly improve the quality of the results. For example, it is
possible to replace the noninformative priors by nested sequences of
priors and hyperpriors~\cite{peixoto_nonparametric_2017} that enhance
our capacity to identify small groups in large networks, more adequately
describe broad degree sequences, and uncover hierarchical modular
structures~\cite{peixoto_hierarchical_2014}. We can also incorporate the
existence of edge covariates~\cite{peixoto_nonparametric_2018}, and the
existence of latent edges for network
reconstruction~\cite{peixoto_reconstructing_2018,peixoto_network_2019,peixoto_latent_2020}. The
method we will describe in the following is suitable for all of these
scenarios, and does not depend on the details of the model
specification. It is able to asymptotically sample from an arbitrary
desired posterior distribution, for which we use a shortcut notation
\begin{equation}
  \pi(\bb) = P(\bb|\A),
\end{equation}
defined up to an unknown normalization constant.

\section{Markov chain Monte Carlo (MCMC)}

The MCMC protocol consists of starting from an arbitrary partition $\bb$
of the network, and then sampling a new partition $\bb'$ from a proposal
distribution
\begin{equation}
  P(\bb'|\bb)
\end{equation}
conditioned on the current partition. The new proposal is accepted with
a Metropolis-Hastings
(MH)~\cite{metropolis_equation_1953,hastings_monte_1970} probability
\begin{equation}
  \min\left(1,\frac{\pi(\bb')P(\bb|\bb')}{\pi(\bb)P(\bb'|\bb)}\right);
\end{equation}
otherwise the move is rejected, and the next state of the chain is the
same as the previous one. Note that to compute the ratio
$\pi(\bb')/\pi(\bb)$ we need to know $\pi(\bb)$ only up to a
normalization constant, which cancels out. The use of this choice
guarantees that the final transition probabilities $T(\bb'|\bb)$, after
considering the acceptance or rejection, fulfills the detailed balance
condition
\begin{equation}
  \pi(\bb)T(\bb'|\bb) = \pi(\bb')T(\bb|\bb').
\end{equation}
When this condition is combined with ergodicity, i.e. the proposals
$P(\bb'|\bb)$ allow the eventual exploration of every possible
partition, and lack of periodicity, i.e. the chain can return to a
previous state in a number of steps that is not necessarily confined to
a multiple of an integer larger than one, then, after a sufficiently
large number of steps the partitions are visited by the chain with a
probability that converges to the desired distribution $\pi(\bb)$.

What determines whether the MCMC method works in practice is the quality
of the proposals $P(\bb'|\bb)$. Clearly, if the proposal happens to be
identical to the target distribution, then the chain mixes in a single
step. But if it were possible to propose partitions directly from the
target distribution, then we would not need MCMC in the first place. In
reality we need to rely on proposals that are far away from the target
distribution, but nevertheless allow the chain to converge to it in a
feasible amount of time. In the following, we describe the conventional
approach of proposing the change of a single node at a time, and then we
introduce moves involving more than one node.

\subsection{Single-node moves}\label{sec:mcmc}

The simplest kind of move proposal that can be done is the one that
involves the change of a single node $i$ from its current group $b_i=r$
to another group $s$, according to a local move proposal $P(s|i,\bb)$,
leading to
\begin{equation}\label{eq:single-prop}
  P(\bb'|\bb)=\sum_iP(i)\prod_j\left[P(b'_j|j,\bb)\right]^{\delta_{ij}}\left[\delta_{b'_i,b_i}\right]^{1-\delta_{ij}},
\end{equation}
with $P(i)$ being the probability of choosing node $i$ to perform such a
move. Note that we do not need to compute this whole expression when
calculating the MH acceptance criterion, as the forward and reverse
proposal must involve the same node, which also means that the
probability $P(i)$ will cancel out, and hence we are allowed to choose
nodes with arbitrary frequencies, as long as they are nonzero to
guarantee ergodicity.

A simple choice for the local move proposal would be to select the
target group $s$ uniformly at random from all $B(\bb)+1$ possibilities,
including a new previously unoccupied group, where $B(\bb)$ is the
number of occupied groups in partition $\bb$, i.e.
\begin{equation}
  P(s|i,\bb) = \frac{1}{B(\bb)+1}.
\end{equation}
Although this is straightforward to implement, it will be inefficient as
soon as the typical number of groups starts to moderately
increase. Suppose, for example, that the typical number of groups is
around $20$ but any given node can belong with nonnegligible
probability to at most $3$ groups. In such a situation $17/20=.85$ of
all move proposals will be rejected, leading to a substantial waste of
effort, which becomes worse as the number of groups increases. We can be
more efficient by proposing smarter moves that are less likely to be
rejected. Following Ref.~\cite{peixoto_efficient_2014}, we choose to
move to existing groups with a probability given by
\begin{equation}\label{eq:smart}
  P_e(s|i,\bb) = \sum_t w_{it}\frac{e_{ts} + \epsilon}{e_t+\epsilon B(\bb)},
\end{equation}
where $w_{it}=\sum_jA_{ji}\delta_{b_j,t}/k_i$ is the fraction of
neighbors of node $i$ that belong to group $t$. The above means that we
inspect the local neighborhood of a node $i$, by sampling one of its
neighbors $j$ at random. We then consider its group membership $b_j=t$
and sample our new group $s$ based on the frequency with which it
connects to nodes of type $t$, which is proportional to total number of
edges between these two groups plus a constant,
i.e. $e_{ts}+\epsilon$. The constant $\epsilon$ must be nonzero to
guarantee ergodicity, such that every possible move can be eventually
chosen (a reasonable choice is simply $\epsilon=1$). This kind of move
tends to concentrate more strongly on more plausible moves, thus
increasing the acceptance ratio, although the extent of this improvement
will depend on the network, and how far the chain has progressed from
its initial state. However, this move proposal still does not completely
guarantee ergodicity because it prevents the occupation of a new group
(which in turn forbids the vacancy of any group, as the MH acceptance
probability becomes zero in this case). We can finally incorporate this
kind of move by augmenting our proposals as
\begin{equation}\label{eq:single}
  P(s|i,\bb) =
  \begin{cases}
  d & \text{ if } s \text{ is a new group,}\\
  (1-d) P_e(s|i,\bb) & \text{ otherwise,}
  \end{cases}
\end{equation}
where $d$ is the probability with which the population of a new group is
attempted.

These smarter move proposals can be performed efficiently by
incorporating only a small amount of extra bookkeeping. If at any given
time we keep a list of edges incident on each group, the we can sample a
new group $s$ from the proposal of Eq.~\ref{eq:single} in time $O(1)$,
provided we can sample a random neighbor also in constant time. The
computation of the MH acceptance probability requires us to visit every
neighbor, and hence can be done in time $O(k_i)$, where $k_i$ is the
degree of the node being moved. The ratio $\pi(\bb')/\pi(\bb)$ can also
be computed in time $O(k_i)$ as well, since the changes in
Eq.~\ref{eq:marginal_likelihood} after the move involve a number of
terms that is at most proportional to the degree of the node $i$. This
means that an entire ``sweep'' of the algorithm, which corresponds to
$N$ single-node proposals, can be performed in time $O(E + N)$ where $E$
is the number of edges. We note, however, that this is not true for all
possible parametrizations of the SBM. For instance, the one used by
Riolo \textit{et al}~\cite{riolo_efficient_2017} requires a number of
updated terms on the order of $O(k_i + B)$ where $B$ is the number of
currently occupied groups. This number can increase algebraically with
$N$, and hence that approach will incur a performance degradation as the
size of the network increases.

\begin{figure}
  \includegraphics[width=\columnwidth]{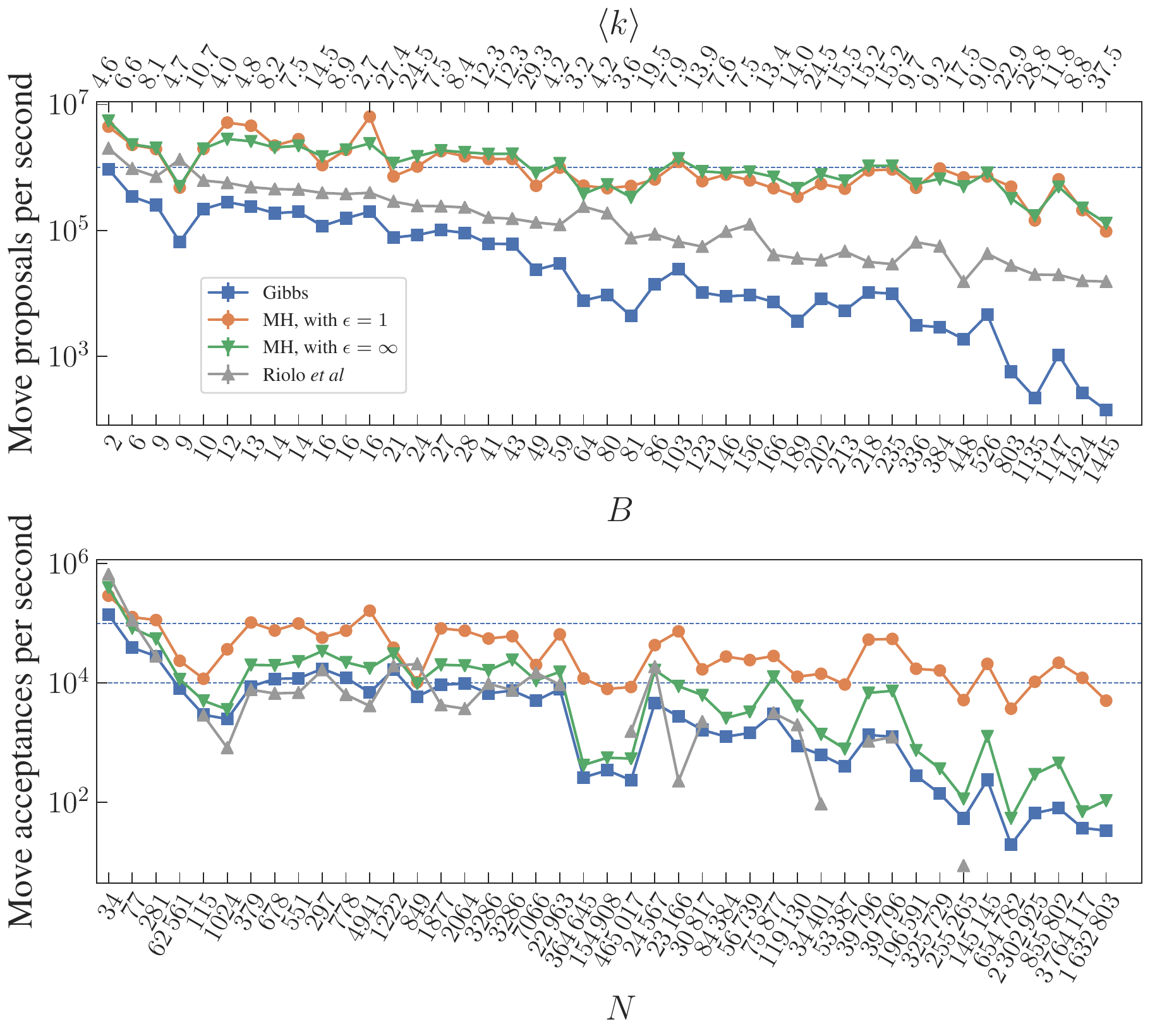}

  \caption{Number of single-node move proposals (top panel) and
  acceptances (bottom panel) per second, for several empirical networks
  gathered from the KONECT repository~\cite{kunegis_konect:_2013}, with
  number of nodes $N$, average degree $\avg{k}$ and occupied groups $B$
  indicated in axis legends. The different curves correspond to the
  single-node moves described in the text, as well as the algorithm by
  Riolo \textit{et al}~\cite{riolo_efficient_2017}, as shown in the
  legend. Missing points indicate a lack of recorded events in the
  allotted time. All algorithms were initialized with the same partition
  obtained with the agglomerative heuristic
  of~\cite{peixoto_efficient_2014}, and run for the same amount of
  time. \label{fig:moves}}
\end{figure}

In Fig.~\ref{fig:moves} we show the achieved number of proposed and
accepted moves per second for a variety of empirical networks collected
from the KONECT repository~\cite{kunegis_konect:_2013}, using an Intel
Xeon Gold 6126 CPU, and with an implementation of the above algorithm
using the C++ programming language~\cite{peixoto_graph-tool_2014}. We
compare the single-node moves above with $\epsilon=1$ and
$\epsilon=\infty$ (which amounts to fully random moves, for which the
extra bookkeeping is disabled), as well as a Gibbs (or ``heat bath'')
sampling version that corresponds to a move proposal given by
\begin{equation}
  P(s|i,\bb) = \frac{\pi(b_1,\dots, b_i=s, \dots, b_N)}{\sum_t\pi(b_1,\dots, b_i=t, \dots, b_N)}.
\end{equation}
Note that this move proposal always has a MH acceptance probability of
$1$, however to implement it we need to compute the probability for the
move of node $i$ to every other group, resulting in a required time
$O(k_iB)$ for every move. We also compare with the approach by Riolo
\textit{et al}~\cite{riolo_efficient_2017} that proposes single-node
moves according to group sizes, using the C implementation made
available by the authors. As we see in Fig.~\ref{fig:moves}, our method
described above achieves millions of proposals per second, which can be
sustained for networks of very large size. The extra bookkeeping
required by the smart move proposals incurs a barely noticeable
slow-down of the rate of proposals. In fact, in certain cases it seems
to speed up the algorithm, which may seem counter-intuitive. This
happens because the proposal in these cases often chooses to move a node
to the same group it is currently in, in which case our implementation
simply skips the move altogether, saving some computation time. Both the
Gibbs sampling and the Riolo \textit{et al} method show a considerable
degradation in performance as the size of the network increases, which
is due to their explicit linear time dependency on the number of groups,
as discussed previously. In terms of the acceptance rate, the smart move
proposals achieve a roughly constant number of accepted moves in the
range of $10^4$ to $10^5$ per second, independent of the size of the
network. When performing fully random moves with $\epsilon=\infty$, the
performance degrades noticeably, due to the problem with a large number
of groups as we described previously. Despite the ideal acceptance rate
with Gibbs sampling, the actual number of acceptances obtained with it
is far lower in most cases even when compared to fully random moves,
which is due entirely to the higher computational cost of the move
proposals. In terms of the rate of move acceptances, the performance of
the Riolo \textit{et al} algorithm is often equivalent to Gibbs
sampling, even though it is based on a MH scheme. This is due to a
combination of the performance degradation that comes from random moves
with the extra time required to compute the likelihood ratio when using
their model parametrization.

Although this kind of performance may seem encouraging at first, it is
not difficult to see that ultimately this kind of approach cannot work
very well. Consider, for example, starting the Markov chain from a
partition consisting of a single group containing all the nodes. The
only possible way forward is for one of the nodes to be placed on a
different group on its own. However, for a majority of networks, such
move will almost always be rejected. By means of a concrete example,
consider the American football network of
Ref.~\cite{girvan_community_2002}, composed of 115 nodes and 613 edges,
which has a clear community structure. For this network, starting from a
single group partition, the easiest node to move to a new group needs to
overcome a posterior probability ratio $\pi(\bb')/\pi(\bb) \approx 6.13
\times 10^{-8}$. This means that such a move will be accepted for the
first time, on average, only after around $10^8$ attempts, and even when
this happens, the next accepted move is overwhelmingly more likely to be
the one that puts this node back into its single group, moving the chain
back to its initial state. The Markov chain, even though it is in theory
ergodic, effectively behaves as if it were not, because it is forced to
go through such states with very low probability, before reaching those
that are more likely. It is clear from this example that the same
problem can happen also when the starting state has more than one group,
for example when two clear communities are merged into one: For these
nodes to be split into two groups, this process must begin by placing
one of the nodes into a new group by itself, which likewise will be
almost always rejected. The same problem occurs also in the reverse
direction: in order for an arbitrary division of a community into two
groups to be merged into one, even if this final state has a much higher
probability, it can only be reached by making one of the groups
progressively smaller, until it is left with a single node, which is
then finally vacated. Even though this last step is likely to be
accepted, the ones preceding it are progressively less probable, and
would almost never be observed in any practical amount of time.

Attempts to alleviate this effective lack of ergodicity include
sophisticated heuristics for initialization of the Markov chain, such as
the agglomerative approach of Ref.~\cite{peixoto_efficient_2014}, which
tries to place the initial position of the chain closer to the most
likely partitions, in the hope these low probability barriers will not
be relevant there. However, this assumption can be easily violated in
situations where posterior distribution ascribes high probabilities to
partitions with different number of groups, and an efficient algorithm
would need to visit them all. In the following, we address this
limitation by allowing the chain to skip over these barriers by
performing the merging and splitting of groups as a single step of the
Markov chain.

\subsection{Merges and splits} \label{sec:mergesplit}

To overcome the limitations of single-node moves, we may introduce moves
where entire groups are merged together. We implement this by selecting
an existing group $r$ with a uniform probability $P(r)=1/B(\bb)$ and
moving all its nodes to a different group $s$ with probability
$P(s|r,\bb)$, yielding a transition proposal
\begin{multline}\label{eq:merge}
  P(\bb'|\bb) = \\
  \sum_{rs}P(r)P(s|r,\bb)\left[\prod_i\left(\delta_{b_i',s}\right)^{\delta_{b_i,r}}\left(\delta_{b_i',b_i}\right)^{1-\delta_{b_i,r}}\right],
\end{multline}
The simplest merge choice is one where the second group is chosen
uniformly at random
\begin{equation}
  P(s|r,\bb) = \frac{1-\delta_{rs}}{B(\bb)-1}.
\end{equation}
However, we have already seen the suboptimal acceptance rate of such
random choices when doing single-node moves, and we should expect the
same problems to occur when doing merges as well. Following the same
line as before, a smarter merge proposal is
\begin{equation}\label{eq:merge-smart}
  P(s|r,\bb) = \frac{1-\delta_{sr}}{n_r}\sum_i\delta_{b_i,r}\frac{P_e(s|i,\bb)}{1-P_e(r|i,\bb)},
\end{equation}
where $P_e(s|i,\bb)$ is given by Eq.~\ref{eq:smart}. This amounts to a
uniformly random selection of a node in group $r$, and using it to
perform the same smart selection of group $s$ as before, but ignoring
the situation where $s=r$, and using that as a merge proposal.

Such merge proposals are straightforward to implement, but in order for
this to incorporated into the MH scheme we need to be able to propose
``split'' moves that go in the reverse direction, i.e. we need to be
able to select a subset of the nodes in group $r$ and move them to a new
group $s$. We denote this subset via a binary vector $\x = \{0,1\}^N$,
where $x_i=1$ if node $i$ is selected to move from group $r$ to $s$,
otherwise $x_i=0$. This leads us to a move proposal that can be written
as
\begin{multline}
  P(\bb'|\bb)=\\
  \sum_{r}P(r)\sum_{\x}P(\x|r,\bb)\left[\prod_i\left(\delta_{b_i',s(\bb)}\right)^{\delta_{x_i,1}}\left(\delta_{b_i',b_i}\right)^{\delta_{x_i,0}}\right],
\end{multline}
where $s(\bb)$ denotes the new group not currently occupied in partition
$\bb$ (we do not differentiate between unoccupied groups).  A simple,
but ultimately naive way of moving forward is to sample $\x$ in two
stages: at first we select the number $m$ of nodes to be moved uniformly
at random in the interval $[1,n_r-1]$ with probability
\begin{equation}
  P(m|r,\bb) = \frac{1}{n_r-1},
\end{equation}
and then sample the $m$ nodes uniformly at random, with a
probability
\begin{equation}
  P(\x|r,\bb,m) = \frac{\delta_{m,\sum_ix_i}}{{n_r \choose m}},
\end{equation}
yielding thus
\begin{equation}
  P(\x|r,\bb) = \sum_{m=1}^{n_r-1}P(\x|r,\bb,m)P(m|r,\bb).
\end{equation}
Although this approach is easy to implement, it does not work in
practice, because such fully random split proposals are almost never
accepted, since the number of good splits, even when they exist, is
exponentially outnumbered by bad ones. Furthermore, the above puts a low
probability for every possible split, which means that it will also
cause good merges to be rejected, as they cannot be easily reversed
under this scheme.

At this point, it is important to realize that finding group splits is
just a smaller scale version of our original problem of finding good
overall partitions of the whole network, so we should not expect it be
accomplished in one fell swoop. This inherent difficulty in proposing
splits makes a merge-split MCMC less straightforward than one might
realize at first. Although we might consider employing one of the many
possible heuristics to find such good splits, we stumble upon two
difficulties:
1. Besides sampling from it, we need also to be able to compute the
probability of the proposal, to decide whether it can be accepted;
2. Given a merge proposal, we need to compute the corresponding split
probability in the reverse direction, without actually sampling from the
split proposal. These requirements limit the kind of schemes we have at
our disposal. Luckily, as it has been shown before by Jain and
Neal~\cite{jain_split-merge_2004, jain_splitting_2007} in the context of
Dirichlet process mixture models, we can overcome these limitations and
perform splits with a much higher acceptance rate by making use of a
split ``staging'' step as an auxiliary variable, as we describe in the
following.

\subsubsection{Auxiliary variables and split staging}

Let us recall the detailed balance condition that needs to be fulfilled
for the Markov chain to converge to the target distribution,
\begin{equation}
  \pi(\bb)T(\bb'|\bb) = \pi(\bb')T(\bb|\bb'),
\end{equation}
where $T(\bb'|\bb)$ is the transition probability, after the rejection
step has been considered. Now, let us consider an augmented version of
the posterior distribution obtained by sampling an arbitrary auxiliary
variable $\alpha$ with probability $P(\alpha|\bb)$. If we use MCMC to
sample from the joint distribution
$P(\alpha,\bb)=P(\alpha|\bb)\pi(\bb)$, then we can marginalize it to
obtain the original distribution $\pi(\bb)=\sum_{\alpha}P(\alpha,\bb)$,
therefore sampling from this augmented space subsumes sampling from the
original one. The usefulness of introducing this auxiliary variable
comes from the fact that we can use it to condition our move proposals,
as we will see. The detailed balance condition in this case reads
\begin{equation}
  P(\alpha|\bb)\pi(\bb)T(\alpha',\bb'|\alpha,\bb) = P(\alpha'|\bb')\pi(\bb')T(\alpha,\bb|\alpha',\bb').
\end{equation}
We can choose the joint transition by first making a transition
$\bb\to\bb'$ and then sampling the auxiliary variable from its
conditional distribution, i.e.
\begin{equation}
  T(\alpha',\bb'|\alpha,\bb) = P(\alpha'|\bb')T(\bb'|\alpha,\bb),
\end{equation}
Then the above detailed balance condition boils down to
\begin{equation}
  \pi(\bb)T(\bb'|\bb,\alpha) = \pi(\bb')T(\bb|\bb',\alpha'),
\end{equation}
which, importantly, is independent of the probability
$P(\alpha|\bb)$. This condition is the same
as the detailed balance for the original system, with the difference
that the transitions are conditioned on arbitrary values of the
auxiliary variable. We can incorporate this into the Metropolis-Hastings
framework by conditioning our proposals $P(\bb'|\bb,\alpha)$, and
accepting them with probability
\begin{equation}
  \min\left(1,\frac{\pi(\bb')P(\bb|\bb',\alpha')}{\pi(\bb)P(\bb'|\bb,\alpha)}\right),
\end{equation}
which will enforce the condition above. The key advantage here is that
we do not need to know how to compute the probability $P(\alpha|\bb)$;
we need only to be able to \emph{sample} from this distribution. In
other words, the transition probabilities themselves can be randomly
sampled, without affecting detailed balance, and without requiring us to
compute their probability. This gives us freedom to implement more
elaborate schemes, and leads us to the notion of \emph{split staging},
where as an auxiliary variable we use a tentative split $\hat{\x}$,
which is again a binary vector with elements $\hat x_i \in \{0,1\}$,
defined for the nodes that belong to the group being split, which is
sampled from some arbitrary distribution (which we will discuss
shortly), and we perform the final split via a single Gibbs sweep from
that initial state, whose probability can be easily computed as
\begin{equation}
  P(\x|r,\bb,\hat\x) = \prod_{i\in V_r}P(x_i|r,\bb, x_1,\dots,x_{i-1},\hat x_{i+1},\dots, \hat x_N),
\end{equation}
where $r$ is the group being split, $V_r$ is the set of nodes that
belong to it, and
\begin{multline}
  P(x_i|r,\bb, x_1,\dots,x_{i-1},\hat x_{i+1},\dots, \hat x_N) =\\
  \frac{\pi\left(\bb'(x_1,\dots,x_{i-1},x_i,\hat x_{i+1},\dots, \hat x_N)\right)}
  {\sum_{x=0}^1\pi\left(\bb'(x_1,\dots,x_{i-1},x,\hat x_{i+1},\dots, \hat x_N)\right)}
\end{multline}
is the probability of moving node $i$ to either $x_i=0$ or $1$ in
sequence, with the shortcut notation $\bb'(\x) =
(b_1'(x_1),\dots,b_N'(x_N))$ given by
\begin{equation}
  b_i'(x_i) =
  \begin{cases}
    s & \text{ if } x_i = 1 \text{ and } b_i = r,\\
    b_i & \text{ otherwise. }
  \end{cases}
\end{equation}
This yields a total split proposal probability
\begin{equation}\label{eq:split}
  P(\bb'|\bb,\hat\x)=\sum_{r}P(r)\sum_{\x}P(\x|r,\bb,\hat\x)\prod_i\delta_{b'_i,b_i'(x_i)},
\end{equation}
which is conditioned on the stage split $\hat\x$, sampled from an
arbitrary distribution. Note that this choice guarantees the strict
reversibility of every possible merge since any split has a strict
nonzero probability, regardless of the staging split $\hat\x$.

We need now only to determine how to sample the staging split $\hat\x$,
but we can proceed in a variety of ways, since we do not need to compute
the resulting probabilities. Here we can incorporate previous knowledge
about what kind of heuristics work better for this problem. As was shown
in Ref.~\cite{peixoto_efficient_2014}, starting from a random division,
and then moving one node at a time, usually yields very bad performance,
as this scheme needs a long time to find the optimal division. This
happens even if the optimal division is very clear, with a very high
posterior probability, since the initial random division effectively
hides it from view, and the MCMC sampling essentially does a random walk
in the configuration space, before it can find the best split. As an
alternative, in Ref.~\cite{peixoto_efficient_2014} it was shown that
agglomerative schemes work significantly better, where one first puts
each node in their own group, then proceeds by merging groups together,
until the desired number of groups is reached. What typically happens in
this scheme is that the intermediary partitions reached amount to
subdivisions of the final partition, thus overcoming the entropic
barriers seen by the random initial division scheme, and thus achieving
the good split in a shorter time. However, we need also to consider a
rather unintuitive property of the dual role of our split proposal,
which functions also as a mechanism to \emph{reverse merge
proposals}. Suppose we are considering an ``easy'' merge of two groups
that represent a very bad local division, i.e. if we had a chance to
merge the groups and split them again, we would be able to find a much
better division. In this situation, since our split proposal would
suggest such a bad split only with a vanishingly small probability, this
would also prevent us from merging the two groups, since the unlikely
reversal would severely penalize the MH acceptance probability. However,
if we also sometimes propose ``bad'' splits, then this would allow such
merges to be accepted. Therefore, counterintuitively, what we need is a
variety of split proposals, that are both good and
bad~\cite{wang_smart-dumb/dumb-smart_2015}. After experimentation, we
determined that the following scheme, which combines a variety of
strategies, works well in a majority of cases. We begin by sampling a
prestaging split $\hat\x^0$ uniformly at random from one of the
following three algorithms:
\begin{enumerate}
\item \textbf{Random split:}
  \begin{enumerate}
  \item We sample the split size $m$ uniformly at random from the interval $[1, n_r-1]$.
  \item We choose $\hat\x^0$ uniformly at random from all splits of size $\sum_i\hat x_i^0=m$.
  \end{enumerate}
\item \textbf{Sequential spreading:}
  \begin{enumerate}
  \item We move all nodes in group $r$ to an empty group $t$.
  \item In random order, we chose the nodes sequentially and move them
    either to group $r$ or group $s$, with probability proportional to
    $\pi(\hat b_1,\dots,\hat b_i,\dots,\hat b_N)$ for a node $i$, with
    $\hat\bb$ being the current working partition, except for the first
    two moves, which are to groups $r$ and $s$, to prevent leaving
    either group empty. In the end, to a node with $\hat b_i=r$ we set
    $\hat x_i^0=0$, and if $\hat b_i=s$, we set $\hat x_i^0=1$.
  \end{enumerate}
\item \textbf{Sequential coalescence:}
  \begin{enumerate}
  \item We move each node $i$ in group $r$ to its own previously empty
        group $t_i$, which in the end all have a single node each.
  \item We proceed like 2(b) above.
  \end{enumerate}
\end{enumerate}
 From the prestage $\hat\x^0$, we obtain $\hat\x$ by performing $M$
 sequential Gibbs sweeps for every node in group $r$ and $s$, where they
 are allowed to move only between these two groups, forbidding moves
 that would leave either group empty. Note that although we could in
 principle compute the probability of proposing the premerge step, this
 becomes intractable after the Gibbs sweeps are performed, since we
 would need to compute the probability of reaching the final state via
 every possible intermediate trajectory. But this is precisely what is
 not needed in this scheme.

 With the stage split in place, the final proposal is obtained by
 performing one more Gibbs sweep as described previously, and it is then
 accepted with probability
\begin{equation}
  \min\left(1,\frac{\pi(\bb')P(\bb|\bb')}{\pi(\bb)P(\bb'|\bb,\hat\x)}\right),
\end{equation}
where the reverse transition $P(\bb|\bb')$ corresponds to a merge
proposal. Likewise a merge can be accepted with probability
\begin{equation}
  \min\left(1,\frac{\pi(\bb')P(\bb|\bb',\hat\x)}{\pi(\bb)P(\bb'|\bb)}\right),
\end{equation}
which means we always need to generate a staging split $\hat\x$ for each
merge candidate, using the algorithm above, to compute the reverse
proposal probability.

As the number $M$ of Gibbs sweeps made during the staging step
increases, the more likely it becomes that the split will be proposed
with a probability proportional to the target distribution $\pi(\bb)$,
which would be optimal. In practice, however, we do not want to choose
this value too large, as it is not worth to spend too much time in a
single Monte Carlo step. Although the optimal value is likely to vary
for each network, we found that a value around $M=10$ offers a good
trade-off between speed and proposal quality in most experiments we
made.

\subsubsection{Joint merge-split moves}

We also consider an additional kind of move that keeps the number of
groups constant, and is composed of a merge of group $r$ into $s$, which
is then split again by moving some nodes from $s$ to back to $r$,
following the same scheme as before. This amounts to a proposal
probability
\begin{multline}\label{eq:merge-split}
  P(\bb'|\bb,\hat\x)=\\
  \sum_{rs}P(r)P(s|r,\bb)\sum_{\hat\bb}\left[\prod_i\left(\delta_{\hat b_i,s}\right)^{\delta_{b_i,r}}\left(\delta_{\hat b_i,b_i}\right)^{1-\delta_{b_i,r}}\right]\times\\
  \sum_{\x}P(\x|s,\hat\bb,\hat\x)\prod_i\delta_{b_i'(x_i),b_i'},
\end{multline}
and the final move is accepted with probability
\begin{equation}\label{eq:merge-split-MH}
  \min\left(1,\frac{\pi(\bb')P(\bb|\bb',\hat\x')}{\pi(\bb)P(\bb'|\bb,\hat\x)}\right).
\end{equation}
This kind of move is able to re-distribute the nodes between two groups,
without going through states with a smaller or lower number of groups,
which we found to improve mixing in circumstances where the number of
groups tends not to vary significantly in the posterior distribution.

\subsection{Algorithm overview and overall complexity}

Even though the split and merge moves are strictly sufficient to
guarantee ergodicity and detailed balance, the mixing time of the Markov
chain is improved if we combine all types of moves considered above. We
do so by introducing the relative move propensities
$\omega_{\text{single}}$, $\omega_{\text{merge}}$,
$\omega_{\text{split}}$ and $\omega_{\text{merge-split}}$, such that,
e.g. the probability of a single-node move is given by
\begin{equation}\label{eq:move-choose}
  \frac{\omega_{\text{single}}}{\omega_{\text{single}} + \omega_{\text{merge}} +
    \omega_{\text{split}} + \omega_{\text{merge-split}}},
\end{equation}
and likewise for the other kinds of moves. At each step we choose one of
the moves above with the corresponding probability, and accept or reject
according to the appropriate MH criterion. We note that for merge and
splits we need to incorporate these move probabilities into the
rejection criterion, e.g. for merge proposals we have
\begin{equation}\label{eq:MH-w}
  \min\left(1,\frac{\pi(\bb')P(\bb|\bb',\hat\x)\,\omega_{\text{split}}}{\pi(\bb)P(\bb'|\bb)\,\omega_{\text{merge}}}\right),
\end{equation}
and likewise for splits. In our experiments, we found it more efficient
to propose single nodes more often, and we have chosen
$\omega_{\text{single}} = N$ and $\omega_{\text{merge}} =
\omega_{\text{split}} = \omega_{\text{merge-split}} = 1$ for our
analysis. It is also possible to determine optimal choices for these
parameters by performing preliminary runs of the algorithm, and then
choosing the values according to the observed acceptance rates.

In summary, the overall structure of the algorithm is as follows:
\begin{enumerate}
  \item At each step, we choose to perform either a single-node, merge, split or
        merge-split move according to Eq.~\ref{eq:move-choose} or the
        analogous for the other moves.
  \item For single node moves, we make a move proposal according to
        Eqs.~\ref{eq:single-prop} and \ref{eq:single}, and we accept it
        according to the MH criterion.
  \item For merges, we make a proposal according to Eqs.~\ref{eq:merge}
        and ~\ref{eq:merge-smart}, and accept it with the MH criterion
        of Eq.~\ref{eq:MH-w}, which involves computing the move reversal
        probability via a split, which is obtained according to
        Eq.~\ref{eq:split}, conditioned on a sampled staged split. The
        staged split is sampled in the same manner as when performing
        split proposals.
  \item For splits, we make a proposal according to Eq.~\ref{eq:split},
        which involves performing a prestage split using either the
        random, sequential spreading or sequential coalescence
        strategies, chosen uniformly at random, followed by $M$ Gibbs
        sweeps to obtain the staged split. From that, we perform the
        final split proposal via a single Gibbs sweep. We accept it
        according to the equivalent MH criterion of Eq.~\ref{eq:MH-w},
        which involves computing the reverse merge proposal probability,
        according to Eq.~\ref{eq:merge}
  \item For merge-splits, we make a proposal according to
        Eq.~\ref{eq:merge-split}, which involves performing both a merge
        and a split, and accept it according to
        Eq.~\ref{eq:merge-split-MH}, which involves a merge-split that
        reverses it.
\end{enumerate}

If the model parametrization of Ref.~\cite{peixoto_nonparametric_2017}
is used, the move of a single node $i$ to another group can be done in
time $O(k_i)$, where $k_i$ is its degree.  Because of this, the time
taken to perform a split of group $r$ is $O[M(n_r + e_r)]$, where $M$ is
the number of Gibbs sweeps used in the split staging, $e_r = \sum_i
k_i\delta_{b_i,r}$ and we need to process every node even if they have
zero degree. Likewise, to merge group $r$ with $s$ we need time $O(n_r +
e_r)$, but since we need to compute the reverse split proposal to obtain
the acceptance probability, we need in fact time $O[M(n_r + n_s + e_r +
  e_s)]$. Since the different move proposals involve a different number
of nodes, their relative probabilities will determine the typical
running time. The longest a merge, split or merge-split move can take is
$O[M(N+E)]$, and hence choosing single-node moves more often with
$\omega_{\text{single}} = O[M(N+E)\times\max(\omega_{\text{merge}},
\omega_{\text{split}},\omega_{\text{merge-split}})]$ means that an
entire sweep of $N$ move proposals can be done in expected linear time
$O(N + E)$. A C++ implementation of the algorithm is freely available as
part of the \texttt{graph-tool} Python
library~\cite{peixoto_graph-tool_2014}.

\section{Performance for empirical networks} \label{sec:empirical}

\begin{figure*}
  \includegraphics[width=\textwidth]{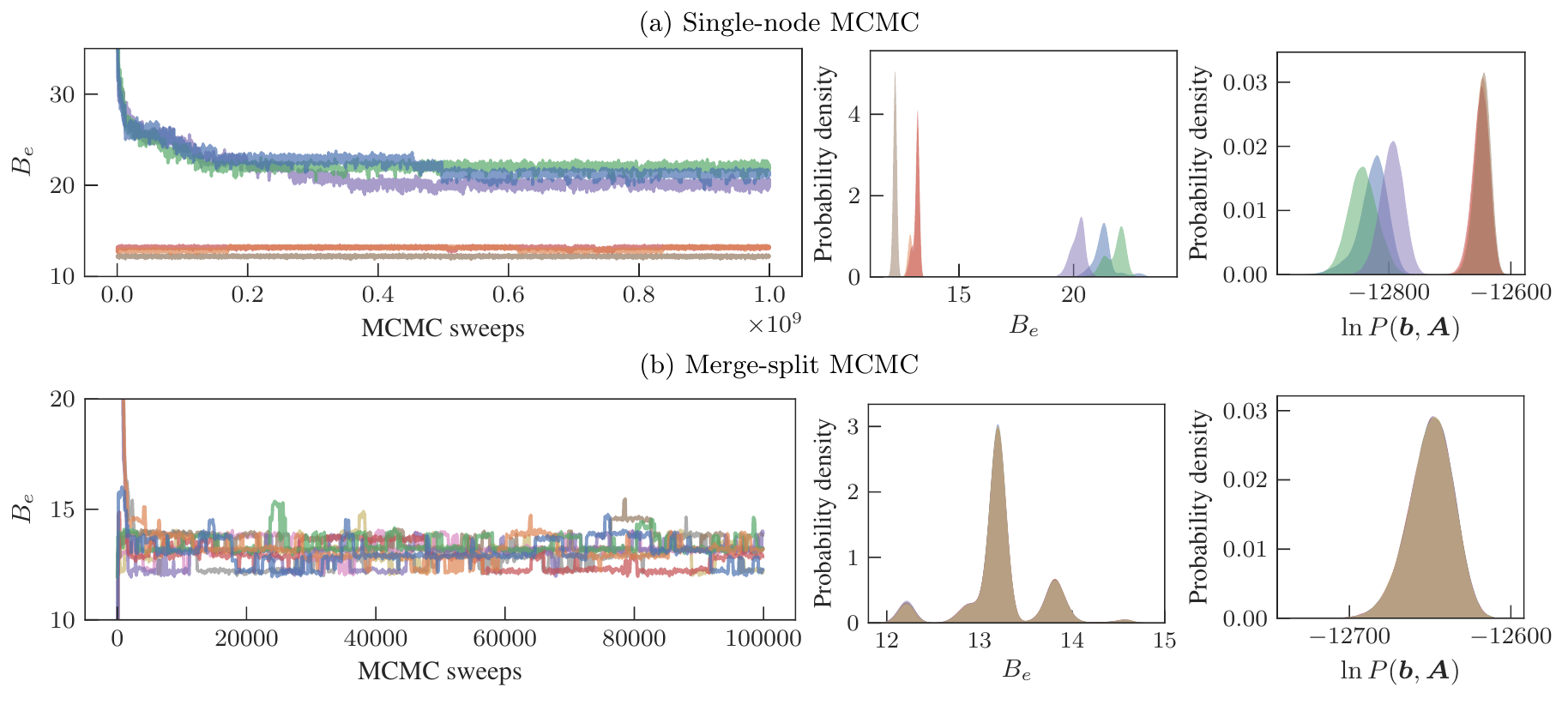}
  \caption{Comparison of runs with the single-node and merge-split MCMC algorithms, for a network of friendships
    between $N=678$ high school students~\cite{moody_peer_2001},
    starting for $B=N$ and $B=1$ states (except for the single-node
    algorithm, where the $B=1$ initialization is replaced with an
    agglomerative heuristic) showing the values of the effective number
    of groups $B_e$ as a function of the number of sweeps (where one
    sweep equals $N$ move proposals), the resulting distribution of
    $B_e$ values and unnormalized posterior probability $P(\bb,\A)$ for
    each run, drawn with different colors.\label{fig:add_health}}
\end{figure*}

In Fig.~\ref{fig:add_health} we show a comparison between the
single-node and merge-split MCMC algorithms for a network of friendships
between $N=678$ high school students~\cite{moody_peer_2001}. We compare
six runs of each algorithm, half of which start from the $B=N$
single-node groups state, and the other half from the single-group $B=1$
state, except for the single-node MCMC runs which would get trivially
trapped in this state, in which case we initialize from an estimate of
the ground state using the agglomerative heuristic described in
Ref.~\cite{peixoto_efficient_2014}. To analyze the convergence of the
chain, we compute the effective number of groups $B_e=\ee^{S}$, where
\begin{equation}
  S = -\sum_r\frac{n_r}{N}\ln \frac{n_r}{N},
\end{equation}
is the entropy of the group membership distribution. As the results
show, the single-node algorithm fails to equilibrate the chain even
after an inordinate number of $10^9$ sweeps, as the different starting
states do not converge to the same values. The initial states with a
large number of groups remain at a relatively larger value, due the
``merge'' barriers we have described previously. More worryingly, even
the improved starting states fail to make the algorithm work, as the
chains clearly remain trapped in different sets of partitions. It is
important to note that individual chains do not show any sign of a lack
of equilibration, as they are seemingly trapped in metastable states,
but they completely fail to accurately represent the target
distribution. It is only when we compare different runs of the algorithm
that we see a problem.

The merge-split MCMC, however, yields a substantially improved
performance, where the diverging starting states converge only after a
few sweeps to the same set of sampled states. The posterior distribution
of effective number of groups $B_e$ and unnormalized posterior
probability $P(\bb,\A)$ are identical between the different runs,
serving as strong evidence for their successful equilibration. The
pronounced multiple peaks in the $B_e$ distribution represent the
multimodality of the posterior distribution that the single-node version
of the algorithm cannot overcome.

\begin{figure}
  \includegraphics[width=\columnwidth]{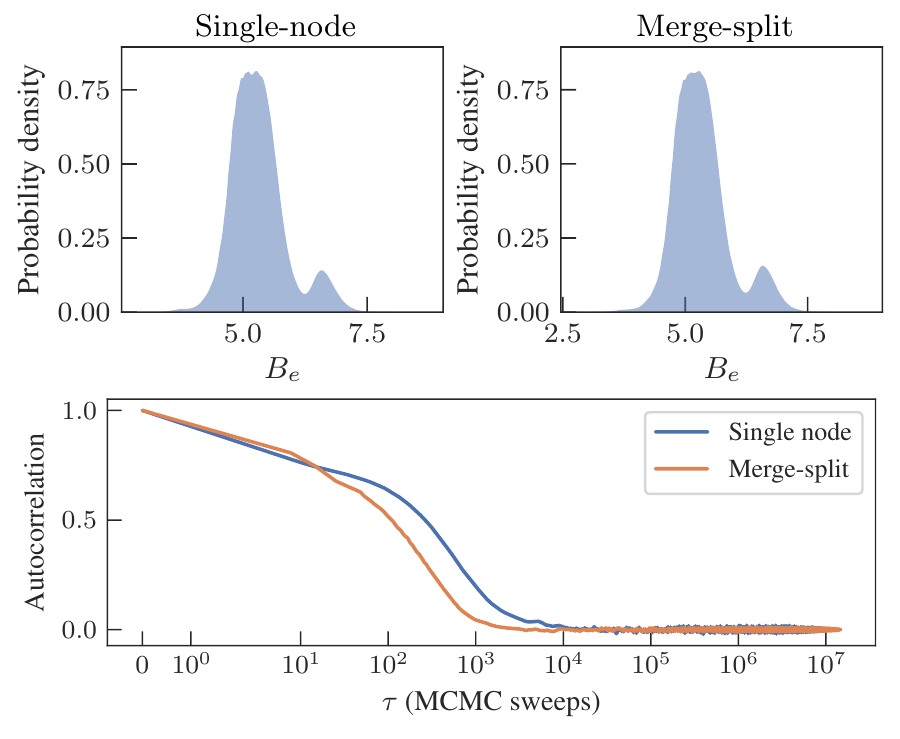}
  \caption{Posterior distribution of the effective number of groups
  $B_e$ obtained with the single-node and merge-split algorithms (top
  panel), as well as the respective autocorrelation functions (bottom
  panel), for the co-appearance network of $N=77$ characters of the
  novel Les Misérables by Victor Hugo~\cite{knuth_stanford_1993}.\label{fig:lesmis}}
\end{figure}

The single-node algorithm can in fact work for networks which are
sufficiently small. For example, if we apply it to the co-appearance
network of $N=77$ characters of the novel Les Misérables by Victor
Hugo~\cite{knuth_stanford_1993}, then we obtain essentially the same
performance with both algorithms, as shown in Fig.~\ref{fig:lesmis}. We
can measure in more detail the mixing time of the Markov chain by
computing the autocorrelation of a surrogate statistics, for example for
$B_e$ we have
\begin{equation}
  \rho(\tau) = \frac{\sum_t[B_e(t) - \avg{B_e}][B_e(t + \tau) - \avg{B_e}]}{\sum_t[B_e(t) - \avg{B_e}]^2},
\end{equation}
where $B_e(t)$ is the value of $B_e$ after $t$ steps of the Markov
chain, and $\avg{B_e}$ is the mean value in the whole range. The value
of $\tau$ for which $\rho(\tau)$ approaches zero gives the
decorrelation time after which we expect samples to be independent
(provided the chain is not trapped in a metastable state). As seen in
Fig.~\ref{fig:lesmis}, the merge-split algorithm de-correlates faster,
but the difference is very small in this example.

\begin{table}
  \begin{tabular}{lrrrr}
    & \multicolumn{2}{c}{Single-node} & \multicolumn{2}{c}{Merge-split}\\
    Network & \multicolumn{1}{c}{$n$} & \multicolumn{1}{c}{ESS} &  \multicolumn{1}{c}{$n$} & \multicolumn{1}{c}{ESS} \\ \hline\hline
    Les Misérables & 5,216,662 & 1,026,851 & 2,984,113 & 2,494,560 \\
    Football & 13,036,736 & 2,532$^*$ & 5,490,028 & 190,267 \\
    High school & 9,907,093 & 7$^*$ & 2,457,300 & 9,945\\ \hline
  \end{tabular}
  \caption{Number of MCMC samples $n$ and effective sample size (ESS)
  obtained with the single-node and merge-split MCMC algorithms, for
  three empirical networks, as described in the text. Values marked with
  $*$ correspond to runs that failed to equilibrate even after a long
  time, and hence the ESS values should not be
  trusted.\label{table:ess}}
\end{table}

\begin{figure}
  \includegraphics[width=\columnwidth]{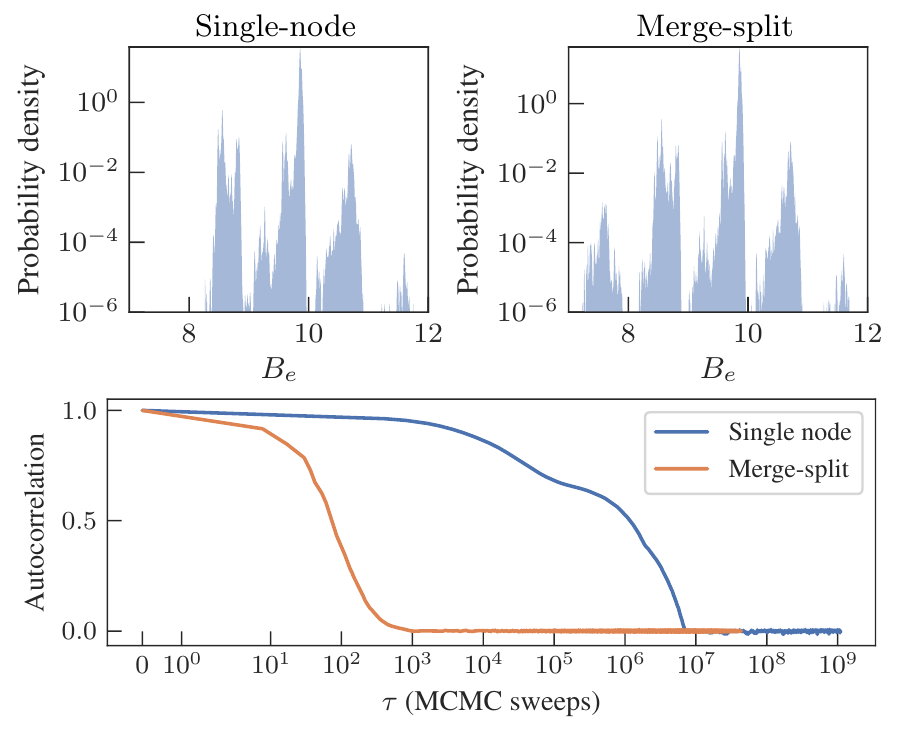}
  \caption{Posterior distribution of the effective number of groups
  $B_e$ obtained with the single-node and merge-split algorithms (top
  panel), as well as the respective autocorrelation functions (bottom
  panel), for the network of American football games between $N=115$
  teams~\cite{girvan_community_2002}.\label{fig:football}}
\end{figure}

\begin{figure}
  \includegraphics[width=\columnwidth]{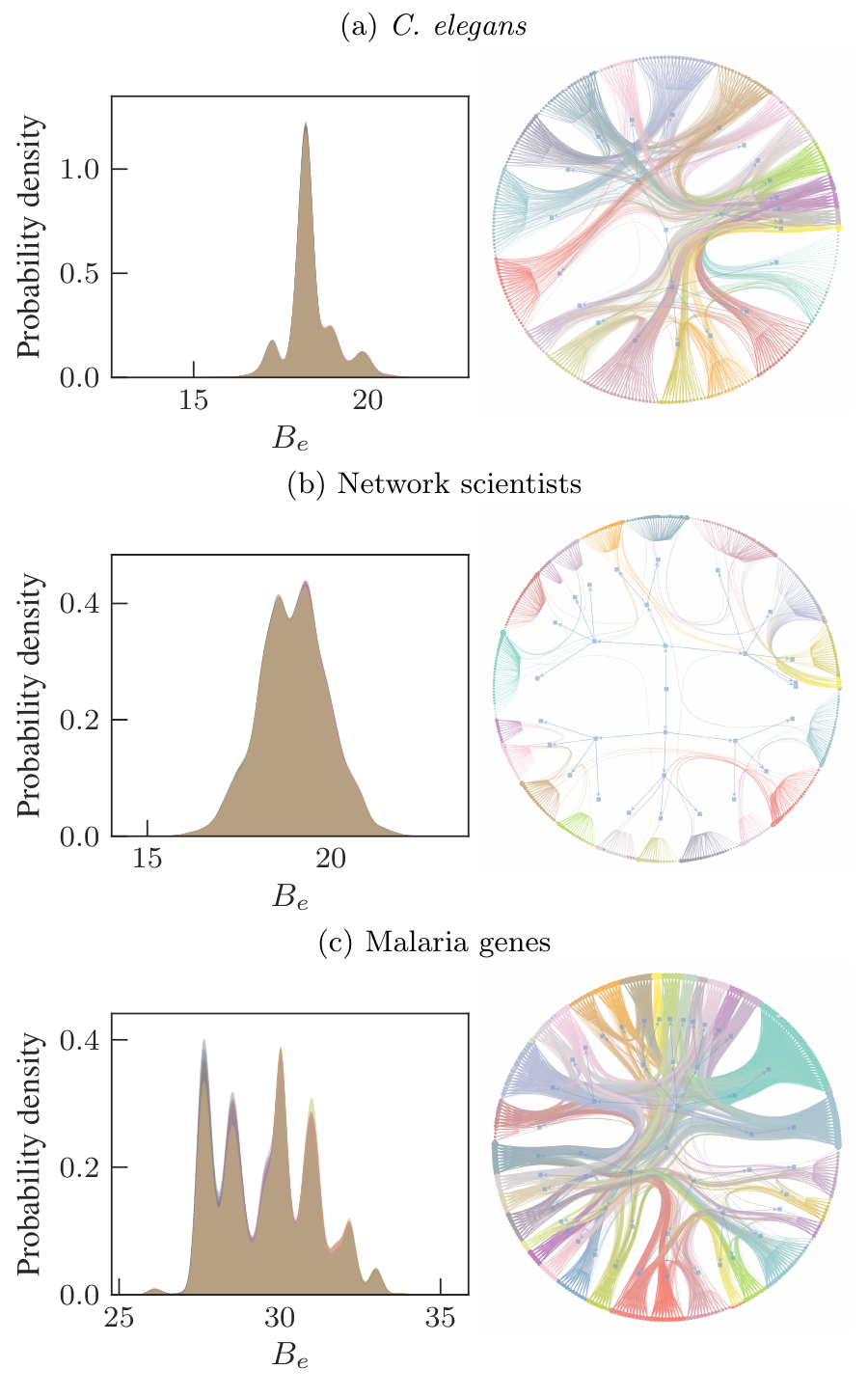}
  \caption{Distribution of the effective number of groups $B_e$ at the
  lowest hierarchical level, obtained for several merge-split MCMC runs,
  starting from different initial conditions (with $B=1$, $B=N$ and an
  approximation of the ground state) drawn with different colors (left)
  and corresponding marginal hierarchical partition (right), for the
  neural network the \emph{C. elegans}
  organism~\cite{white_structure_1986}, the co-authorship network of
  network scientists~\cite{newman_finding_2006}, and the network of
  recombinant antigen genes from the human malaria parasite
  \emph{P. falciparum}~\cite{larremore_network_2013}. In all cases, the
  different distributions coincide, indicating convergence to the target
  distribution.\label{fig:nested}}
\end{figure}

A related quantity that is useful to determine the actual performance of
the MCMC is the effective sample size (ESS) defined as
\begin{equation}
  \text{ESS} = \frac{n}{1+\frac{1}{2}\sum_{\tau=1}^{\infty}\rho(\tau)},
\end{equation}
where $n$ is the total number of samples taken from the MCMC run, and
$\rho(\tau)$ is the autocorrelation computed over only those samples.
The ESS is interpreted as the effective number of uncorrelated samples
taken during the run of the algorithm (provided it has equilibrated). If
the chain is ideal, then we should have $\text{ESS}(n)\approx n$, but
this is often not the case. In Table~\ref{table:ess} we show the number
of samples and the ESS values for the three networks considered
previously, where the samples were gathered after several MCMC
sweeps. For the Les Misérables network, the ESS with the merge-split
algorithm is close to ideal, although the single-node version is still
quite usable. For both the high school and football networks we see a
large difference in the usefulness of both algorithms. In fact, we had
already seen that for the high school network the single-node MCMC
simply fails to converge, meaning that the actual ESS for that case
should be in fact zero. For the football network we see in
Fig.~\ref{fig:football} the autocorrelation obtained with both
algorithms, as well as the posterior $B_e$ distribution. Not only is the
decorrelation four orders of magnitude slower with the single-node
algorithm, but also it in fact fails to correctly sample from all modes
of the distribution, omitting values in the range $B_e \in [7,8]$, even
though they do not have a negligible posterior probability. Therefore,
even though this network is not very large, a detailed sampling from the
posterior distribution using the single-node algorithm is already quite
demanding. For network sizes that are slightly larger, such as the high
school network, it breaks down completely, leaving the merge-split
version as the only viable alternative.

\section{Hierarchical partitions}\label{sec:nested}

\begin{figure}
  \includegraphics[width=\columnwidth]{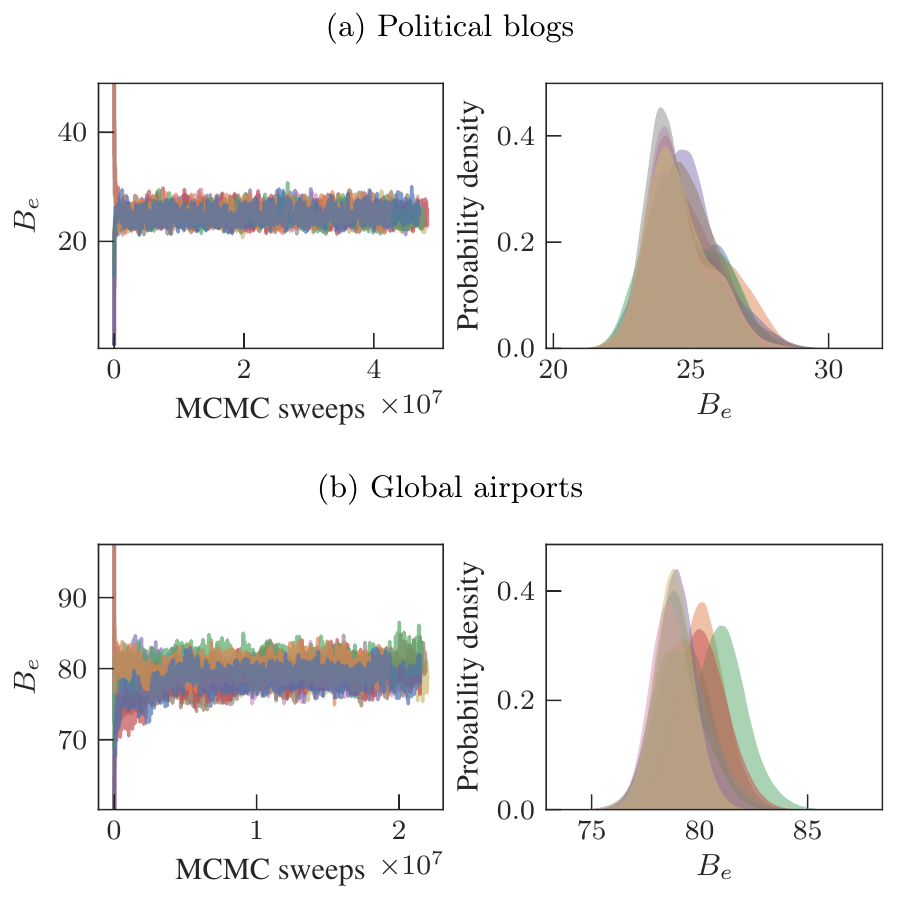}
  \caption{Effective number of groups $B_e$ at the lowest level during
  several long runs of the merge-split MCMC algorithm using the
  hierarchical SBM, drawn with different colors, and corresponding
  distribution, for a network of political
  blogs~\cite{adamic_political_2005} and flights between airports from
  \url{https://openflights.org}. In this case, discrepancies between the
  different distributions are still noticeable after long runs,
  indicating a lack of convergence to the target
  distribution.\label{fig:hard}}
\end{figure}

The nested
SBM~\cite{peixoto_hierarchical_2014,peixoto_nonparametric_2017} consists
of a hierarchical Bayesian model, where the parameters of the SBM are
interpreted as a multigraph, which are then sampled from another SBM,
and so on recursively, forming a nested hierarchy. The nodes at level
$l+1$ are the (nonempty) group labels at level $l$, and we have a set of
partitions $\{\bb_1,\dots, \bb_L\}$, sampled from the posterior
distribution
\begin{equation}
  P(\bb_1,\dots,\bb_L|\A) = \frac{P(\A|\bb_1,\dots,\bb_L)P(\bb_1,\dots,\bb_L)}{P(\A)},
\end{equation}
where
\begin{multline}
  P(\A|\bb_1,\dots,\bb_L)P(\bb_1,\dots,\bb_L) \\
  = P(\A|\bm{e}_1,\bb_1)\prod_{l=1}^{L}P(\bm{e}_l|\bm{e}_{l+1},\bb_{l+1})P(\bb_l|\bb_{l-1}),
\end{multline}
is the marginal likelihood (we refer to
Ref.~\cite{peixoto_nonparametric_2017} for more details).

For this model, we can employ the same MCMC described above, by sampling
from the conditional posterior distribution at each hierarchy level $l$,
\begin{multline}\label{eq:cond_hsbm}
  P(\bb_l|\bb_1,\dots,\bb_{l-1},\bb_{l+1},\dots,\bb_L, \A)\\
  \begin{aligned}
    &= \frac{P(\bb_1,\dots,\bb_L|\A)}{\sum_{\bb'_l}P(\bb_1,\dots,\bb'_l,\dots,\bb_L|\A)}.
  \end{aligned}
\end{multline}
This means we can proceed at each step in the MCMC by choosing a level
$l$ uniformly at random, and setting $\pi(\bb_l) =
P(\bb_l|\bb_1,\dots,\bb_{l-1},\bb_{l+1},\dots,\bb_L, \A)$ as our target
distribution, and making move proposals the exact same way as before, by
performing either a single-node, merge, split, or merge-split move. The
only special consideration needed is when a new group is added at a
given level, a node needs to be added to the level above, together with
its group membership, and possibly so on recursively. This kind of new
group addition needs to be able to reverse the disappearance of a group,
together with its node at the next level, and so on. The way we proceed
is to keep track of unoccupied groups and their own nodes in the
hierarchy, even though they do not contribute to the posterior
probability, with the only purpose of being able to reverse group
vacancies and to propose placement of new groups. This bookkeeping
needs only to be virtual, since at any given time an unoccupied group
can belong to either one of the $B$ occupied groups at the level above
or an unoccupied one, with equal probability. Therefore, whenever a new
group is occupied, its own group membership at level $l$ is chosen
uniformly at random with probability
\begin{equation}
  \frac{1}{B(\bb_l)+1},
\end{equation}
and so on recursively for the higher levels if a new group is chosen.

In Fig.~\ref{fig:nested} we see the above algorithm employed to sample
from the hierarchical partitions of a few empirical networks, in
particular the neural network the \emph{C. elegans}
organism~\cite{white_structure_1986}, the co-authorship network of
network scientists~\cite{newman_finding_2006}, and the network of
recombinant antigen genes from the human malaria parasite
\emph{P. falciparum}~\cite{larremore_network_2013}. For these results we
considered several chains, where the lowest level of the hierarchy was
initialized both at $B=1$ and $B=N$, and in all cases they converged to the
same overall distribution of hierarchical partitions.

The convincing equilibration results we observe for the examples above
are not universal, and for some data the merge-split algorithm is
seemingly not sufficient to fully equilibrate the chain, at least not in
a short amount of time. Some examples of this are shown in
Fig.~\ref{fig:hard} for a network of $N=1,222$ political
blogs~\cite{adamic_political_2005} and flights between $N=3,286$
airports from \url{https://openflights.org}. Although the scheme
succeeds in quickly converging to the same range of partitions, with no
obvious memory of the starting state, the values sampled still show
noticeable discrepancy between the runs, even after a long time. This
long equilibration time is not related to the hierarchical model, and
also occurs when using the ``flat'' model for these examples. The
behavior in this case seems different from the metastable trapping
observed with the single-node scheme, and is probably due to a different
kind of multimodality that the merge and splits are not able to
efficiently evade. This behavior is likely to be an indication of a bad
quality of fit of the SBM for these networks, and requires a strategy
different from the merging and splitting of groups to be effectively
addressed.

\section{Conclusion}\label{sec:conc}

We have showed how merge and split moves can be introduced into MCMC
schemes that sample network partitions from a posterior distribution,
thereby significantly improving the mixing time of the Markov chain in a
variety of empirically relevant scenarios. We have also demonstrated how
the often employed approach of moving the group membership of a single
node at a time is systematically insufficient to characterize the
posterior distribution of group assignments, due to its tendency of
getting trapped in metastable states.

We have extended the algorithm also to the case of hierarchical
partitions, allowing us to sample nested hierarchical clusterings with
the same degree of efficiency. The method developed does not depend on
details of the underlying model, and can be easily extended to other
variations as well.

When investigating the performance of the merge-split scheme in a
variety of empirical networks we also encountered situations where,
although it is far less susceptible to getting trapped as the single
node scheme, it may still need very long runs to sufficiently explore
the posterior landscape of partitions. A better understanding of the
reasons behind the sampling difficulty in these cases may prove
important to a full characterization of the modular network structure in
empirical settings, and in particular to determine the quality of fit of
network models used for community detection. The latter is because this
kind of roughness of the posterior distribution is not expected if the
network is truly sampled from the assumed generative model, and may
point to a unsuitability of the model for that particular data. We leave
this line of inquiry for future work.

\bibliography{bib}

\begin{thebibliography}{33}%
\makeatletter
\providecommand \@ifxundefined [1]{%
 \@ifx{#1\undefined}
}%
\providecommand \@ifnum [1]{%
 \ifnum #1\expandafter \@firstoftwo
 \else \expandafter \@secondoftwo
 \fi
}%
\providecommand \@ifx [1]{%
 \ifx #1\expandafter \@firstoftwo
 \else \expandafter \@secondoftwo
 \fi
}%
\providecommand \natexlab [1]{#1}%
\providecommand \enquote  [1]{``#1''}%
\providecommand \bibnamefont  [1]{#1}%
\providecommand \bibfnamefont [1]{#1}%
\providecommand \citenamefont [1]{#1}%
\providecommand \href@noop [0]{\@secondoftwo}%
\providecommand \href [0]{\begingroup \@sanitize@url \@href}%
\providecommand \@href[1]{\@@startlink{#1}\@@href}%
\providecommand \@@href[1]{\endgroup#1\@@endlink}%
\providecommand \@sanitize@url [0]{\catcode `\\12\catcode `\$12\catcode
  `\&12\catcode `\#12\catcode `\^12\catcode `\_12\catcode `\%12\relax}%
\providecommand \@@startlink[1]{}%
\providecommand \@@endlink[0]{}%
\providecommand \url  [0]{\begingroup\@sanitize@url \@url }%
\providecommand \@url [1]{\endgroup\@href {#1}{\urlprefix }}%
\providecommand \urlprefix  [0]{URL }%
\providecommand \Eprint [0]{\href }%
\providecommand \doibase [0]{http://dx.doi.org/}%
\providecommand \selectlanguage [0]{\@gobble}%
\providecommand \bibinfo  [0]{\@secondoftwo}%
\providecommand \bibfield  [0]{\@secondoftwo}%
\providecommand \translation [1]{[#1]}%
\providecommand \BibitemOpen [0]{}%
\providecommand \bibitemStop [0]{}%
\providecommand \bibitemNoStop [0]{.\EOS\space}%
\providecommand \EOS [0]{\spacefactor3000\relax}%
\providecommand \BibitemShut  [1]{\csname bibitem#1\endcsname}%
\let\auto@bib@innerbib\@empty
\bibitem [{\citenamefont {Fortunato}(2010)}]{fortunato_community_2010}%
  \BibitemOpen
  \bibfield  {author} {\bibinfo {author} {\bibfnamefont {Santo}\ \bibnamefont
  {Fortunato}},\ }\bibfield  {title} {\enquote {\bibinfo {title} {Community
  detection in graphs},}\ }\href {\doibase 16/j.physrep.2009.11.002} {\bibfield
   {journal} {\bibinfo  {journal} {Physics Reports}\ }\textbf {\bibinfo
  {volume} {486}},\ \bibinfo {pages} {75--174} (\bibinfo {year}
  {2010})}\BibitemShut {NoStop}%
\bibitem [{\citenamefont
  {Peixoto}(2019{\natexlab{a}})}]{peixoto_bayesian_2019}%
  \BibitemOpen
  \bibfield  {author} {\bibinfo {author} {\bibfnamefont {Tiago~P.}\
  \bibnamefont {Peixoto}},\ }\bibfield  {title} {{\selectlanguage
  {english}\enquote {\bibinfo {title} {Bayesian {Stochastic}
  {Blockmodeling}},}\ }}in\ \href {\doibase 10.1002/9781119483298.ch11}
  {{\selectlanguage {english}\emph {\bibinfo {booktitle} {Advances in {Network}
  {Clustering} and {Blockmodeling}}}}}\ (\bibinfo  {publisher} {John Wiley \&
  Sons, Ltd},\ \bibinfo {year} {2019})\ pp.\ \bibinfo {pages}
  {289--332}\BibitemShut {NoStop}%
\bibitem [{\citenamefont {Guimerà}\ \emph {et~al.}(2004)\citenamefont
  {Guimerà}, \citenamefont {Sales-Pardo},\ and\ \citenamefont
  {Amaral}}]{guimera_modularity_2004}%
  \BibitemOpen
  \bibfield  {author} {\bibinfo {author} {\bibfnamefont {Roger}\ \bibnamefont
  {Guimerà}}, \bibinfo {author} {\bibfnamefont {Marta}\ \bibnamefont
  {Sales-Pardo}}, \ and\ \bibinfo {author} {\bibfnamefont {Luís A.~Nunes}\
  \bibnamefont {Amaral}},\ }\bibfield  {title} {\enquote {\bibinfo {title}
  {Modularity from fluctuations in random graphs and complex networks},}\
  }\href {\doibase 10.1103/PhysRevE.70.025101} {\bibfield  {journal} {\bibinfo
  {journal} {Physical Review E}\ }\textbf {\bibinfo {volume} {70}},\ \bibinfo
  {pages} {025101} (\bibinfo {year} {2004})}\BibitemShut {NoStop}%
\bibitem [{\citenamefont {Bagrow}(2012)}]{bagrow_communities_2012}%
  \BibitemOpen
  \bibfield  {author} {\bibinfo {author} {\bibfnamefont {James~P.}\
  \bibnamefont {Bagrow}},\ }\bibfield  {title} {\enquote {\bibinfo {title}
  {Communities and bottlenecks: {Trees} and treelike networks have high
  modularity},}\ }\href {\doibase 10.1103/PhysRevE.85.066118} {\bibfield
  {journal} {\bibinfo  {journal} {Physical Review E}\ }\textbf {\bibinfo
  {volume} {85}},\ \bibinfo {pages} {066118} (\bibinfo {year}
  {2012})}\BibitemShut {NoStop}%
\bibitem [{\citenamefont {Fortunato}\ and\ \citenamefont
  {Barthélemy}(2007)}]{fortunato_resolution_2007}%
  \BibitemOpen
  \bibfield  {author} {\bibinfo {author} {\bibfnamefont {Santo}\ \bibnamefont
  {Fortunato}}\ and\ \bibinfo {author} {\bibfnamefont {Marc}\ \bibnamefont
  {Barthélemy}},\ }\bibfield  {title} {{\selectlanguage {english}\enquote
  {\bibinfo {title} {Resolution limit in community detection},}\ }}\href
  {\doibase 10.1073/pnas.0605965104} {\bibfield  {journal} {\bibinfo  {journal}
  {Proceedings of the National Academy of Sciences}\ }\textbf {\bibinfo
  {volume} {104}},\ \bibinfo {pages} {36--41} (\bibinfo {year}
  {2007})}\BibitemShut {NoStop}%
\bibitem [{\citenamefont {Holland}\ \emph {et~al.}(1983)\citenamefont
  {Holland}, \citenamefont {Laskey},\ and\ \citenamefont
  {Leinhardt}}]{holland_stochastic_1983}%
  \BibitemOpen
  \bibfield  {author} {\bibinfo {author} {\bibfnamefont {Paul~W.}\ \bibnamefont
  {Holland}}, \bibinfo {author} {\bibfnamefont {Kathryn~Blackmond}\
  \bibnamefont {Laskey}}, \ and\ \bibinfo {author} {\bibfnamefont {Samuel}\
  \bibnamefont {Leinhardt}},\ }\bibfield  {title} {\enquote {\bibinfo {title}
  {Stochastic blockmodels: {First} steps},}\ }\href {\doibase
  16/0378-8733(83)90021-7} {\bibfield  {journal} {\bibinfo  {journal} {Social
  Networks}\ }\textbf {\bibinfo {volume} {5}},\ \bibinfo {pages} {109--137}
  (\bibinfo {year} {1983})}\BibitemShut {NoStop}%
\bibitem [{\citenamefont {Karrer}\ and\ \citenamefont
  {Newman}(2011)}]{karrer_stochastic_2011}%
  \BibitemOpen
  \bibfield  {author} {\bibinfo {author} {\bibfnamefont {Brian}\ \bibnamefont
  {Karrer}}\ and\ \bibinfo {author} {\bibfnamefont {M.~E.~J.}\ \bibnamefont
  {Newman}},\ }\bibfield  {title} {\enquote {\bibinfo {title} {Stochastic
  blockmodels and community structure in networks},}\ }\href {\doibase
  10.1103/PhysRevE.83.016107} {\bibfield  {journal} {\bibinfo  {journal}
  {Physical Review E}\ }\textbf {\bibinfo {volume} {83}},\ \bibinfo {pages}
  {016107} (\bibinfo {year} {2011})}\BibitemShut {NoStop}%
\bibitem [{\citenamefont {Decelle}\ \emph {et~al.}(2011)\citenamefont
  {Decelle}, \citenamefont {Krzakala}, \citenamefont {Moore},\ and\
  \citenamefont {Zdeborová}}]{decelle_asymptotic_2011}%
  \BibitemOpen
  \bibfield  {author} {\bibinfo {author} {\bibfnamefont {Aurelien}\
  \bibnamefont {Decelle}}, \bibinfo {author} {\bibfnamefont {Florent}\
  \bibnamefont {Krzakala}}, \bibinfo {author} {\bibfnamefont {Cristopher}\
  \bibnamefont {Moore}}, \ and\ \bibinfo {author} {\bibfnamefont {Lenka}\
  \bibnamefont {Zdeborová}},\ }\bibfield  {title} {\enquote {\bibinfo {title}
  {Asymptotic analysis of the stochastic block model for modular networks and
  its algorithmic applications},}\ }\href {\doibase 10.1103/PhysRevE.84.066106}
  {\bibfield  {journal} {\bibinfo  {journal} {Physical Review E}\ }\textbf
  {\bibinfo {volume} {84}},\ \bibinfo {pages} {066106} (\bibinfo {year}
  {2011})}\BibitemShut {NoStop}%
\bibitem [{\citenamefont {Snijders}\ and\ \citenamefont
  {Nowicki}(1997)}]{snijders_estimation_1997}%
  \BibitemOpen
  \bibfield  {author} {\bibinfo {author} {\bibfnamefont {Tom A.~B.}\
  \bibnamefont {Snijders}}\ and\ \bibinfo {author} {\bibfnamefont {Krzysztof}\
  \bibnamefont {Nowicki}},\ }\bibfield  {title} {{\selectlanguage
  {english}\enquote {\bibinfo {title} {Estimation and {Prediction} for
  {Stochastic} {Blockmodels} for {Graphs} with {Latent} {Block} {Structure}},}\
  }}\href {\doibase 10.1007/s003579900004} {\bibfield  {journal} {\bibinfo
  {journal} {Journal of Classification}\ }\textbf {\bibinfo {volume} {14}},\
  \bibinfo {pages} {75--100} (\bibinfo {year} {1997})}\BibitemShut {NoStop}%
\bibitem [{\citenamefont {Guimerà}\ and\ \citenamefont
  {Sales-Pardo}(2009)}]{guimera_missing_2009}%
  \BibitemOpen
  \bibfield  {author} {\bibinfo {author} {\bibfnamefont {Roger}\ \bibnamefont
  {Guimerà}}\ and\ \bibinfo {author} {\bibfnamefont {Marta}\ \bibnamefont
  {Sales-Pardo}},\ }\bibfield  {title} {\enquote {\bibinfo {title} {Missing and
  spurious interactions and the reconstruction of complex networks},}\ }\href
  {\doibase 10.1073/pnas.0908366106} {\bibfield  {journal} {\bibinfo  {journal}
  {Proceedings of the National Academy of Sciences}\ }\textbf {\bibinfo
  {volume} {106}},\ \bibinfo {pages} {22073 --22078} (\bibinfo {year}
  {2009})}\BibitemShut {NoStop}%
\bibitem [{\citenamefont
  {Peixoto}(2014{\natexlab{a}})}]{peixoto_efficient_2014}%
  \BibitemOpen
  \bibfield  {author} {\bibinfo {author} {\bibfnamefont {Tiago~P.}\
  \bibnamefont {Peixoto}},\ }\bibfield  {title} {\enquote {\bibinfo {title}
  {Efficient {Monte} {Carlo} and greedy heuristic for the inference of
  stochastic block models},}\ }\href {\doibase 10.1103/PhysRevE.89.012804}
  {\bibfield  {journal} {\bibinfo  {journal} {Physical Review E}\ }\textbf
  {\bibinfo {volume} {89}},\ \bibinfo {pages} {012804} (\bibinfo {year}
  {2014}{\natexlab{a}})}\BibitemShut {NoStop}%
\bibitem [{\citenamefont {Newman}\ and\ \citenamefont
  {Reinert}(2016)}]{newman_estimating_2016}%
  \BibitemOpen
  \bibfield  {author} {\bibinfo {author} {\bibfnamefont {M.~E.~J.}\
  \bibnamefont {Newman}}\ and\ \bibinfo {author} {\bibfnamefont {Gesine}\
  \bibnamefont {Reinert}},\ }\bibfield  {title} {\enquote {\bibinfo {title}
  {Estimating the {Number} of {Communities} in a {Network}},}\ }\href {\doibase
  10.1103/PhysRevLett.117.078301} {\bibfield  {journal} {\bibinfo  {journal}
  {Physical Review Letters}\ }\textbf {\bibinfo {volume} {117}},\ \bibinfo
  {pages} {078301} (\bibinfo {year} {2016})}\BibitemShut {NoStop}%
\bibitem [{\citenamefont {Riolo}\ \emph {et~al.}(2017)\citenamefont {Riolo},
  \citenamefont {Cantwell}, \citenamefont {Reinert},\ and\ \citenamefont
  {Newman}}]{riolo_efficient_2017}%
  \BibitemOpen
  \bibfield  {author} {\bibinfo {author} {\bibfnamefont {Maria~A.}\
  \bibnamefont {Riolo}}, \bibinfo {author} {\bibfnamefont {George~T.}\
  \bibnamefont {Cantwell}}, \bibinfo {author} {\bibfnamefont {Gesine}\
  \bibnamefont {Reinert}}, \ and\ \bibinfo {author} {\bibfnamefont {M.~E.~J.}\
  \bibnamefont {Newman}},\ }\bibfield  {title} {\enquote {\bibinfo {title}
  {Efficient method for estimating the number of communities in a network},}\
  }\href {\doibase 10.1103/PhysRevE.96.032310} {\bibfield  {journal} {\bibinfo
  {journal} {Physical Review E}\ }\textbf {\bibinfo {volume} {96}},\ \bibinfo
  {pages} {032310} (\bibinfo {year} {2017})}\BibitemShut {NoStop}%
\bibitem [{\citenamefont {Peixoto}(2017)}]{peixoto_nonparametric_2017}%
  \BibitemOpen
  \bibfield  {author} {\bibinfo {author} {\bibfnamefont {Tiago~P.}\
  \bibnamefont {Peixoto}},\ }\bibfield  {title} {\enquote {\bibinfo {title}
  {Nonparametric {Bayesian} inference of the microcanonical stochastic block
  model},}\ }\href {\doibase 10.1103/PhysRevE.95.012317} {\bibfield  {journal}
  {\bibinfo  {journal} {Physical Review E}\ }\textbf {\bibinfo {volume} {95}},\
  \bibinfo {pages} {012317} (\bibinfo {year} {2017})}\BibitemShut {NoStop}%
\bibitem [{\citenamefont
  {Peixoto}(2014{\natexlab{b}})}]{peixoto_hierarchical_2014}%
  \BibitemOpen
  \bibfield  {author} {\bibinfo {author} {\bibfnamefont {Tiago~P.}\
  \bibnamefont {Peixoto}},\ }\bibfield  {title} {\enquote {\bibinfo {title}
  {Hierarchical {Block} {Structures} and {High}-{Resolution} {Model}
  {Selection} in {Large} {Networks}},}\ }\href {\doibase
  10.1103/PhysRevX.4.011047} {\bibfield  {journal} {\bibinfo  {journal}
  {Physical Review X}\ }\textbf {\bibinfo {volume} {4}},\ \bibinfo {pages}
  {011047} (\bibinfo {year} {2014}{\natexlab{b}})}\BibitemShut {NoStop}%
\bibitem [{\citenamefont
  {Peixoto}(2018{\natexlab{a}})}]{peixoto_nonparametric_2018}%
  \BibitemOpen
  \bibfield  {author} {\bibinfo {author} {\bibfnamefont {Tiago~P.}\
  \bibnamefont {Peixoto}},\ }\bibfield  {title} {\enquote {\bibinfo {title}
  {Nonparametric weighted stochastic block models},}\ }\href {\doibase
  10.1103/PhysRevE.97.012306} {\bibfield  {journal} {\bibinfo  {journal}
  {Physical Review E}\ }\textbf {\bibinfo {volume} {97}},\ \bibinfo {pages}
  {012306} (\bibinfo {year} {2018}{\natexlab{a}})}\BibitemShut {NoStop}%
\bibitem [{\citenamefont
  {Peixoto}(2018{\natexlab{b}})}]{peixoto_reconstructing_2018}%
  \BibitemOpen
  \bibfield  {author} {\bibinfo {author} {\bibfnamefont {Tiago~P.}\
  \bibnamefont {Peixoto}},\ }\bibfield  {title} {\enquote {\bibinfo {title}
  {Reconstructing {Networks} with {Unknown} and {Heterogeneous} {Errors}},}\
  }\href {\doibase 10.1103/PhysRevX.8.041011} {\bibfield  {journal} {\bibinfo
  {journal} {Physical Review X}\ }\textbf {\bibinfo {volume} {8}},\ \bibinfo
  {pages} {041011} (\bibinfo {year} {2018}{\natexlab{b}})}\BibitemShut
  {NoStop}%
\bibitem [{\citenamefont {Peixoto}(2019{\natexlab{b}})}]{peixoto_network_2019}%
  \BibitemOpen
  \bibfield  {author} {\bibinfo {author} {\bibfnamefont {Tiago~P.}\
  \bibnamefont {Peixoto}},\ }\bibfield  {title} {\enquote {\bibinfo {title}
  {Network {Reconstruction} and {Community} {Detection} from {Dynamics}},}\
  }\href {\doibase 10.1103/PhysRevLett.123.128301} {\bibfield  {journal}
  {\bibinfo  {journal} {Physical Review Letters}\ }\textbf {\bibinfo {volume}
  {123}},\ \bibinfo {pages} {128301} (\bibinfo {year}
  {2019}{\natexlab{b}})}\BibitemShut {NoStop}%
\bibitem [{\citenamefont {Peixoto}(2020)}]{peixoto_latent_2020}%
  \BibitemOpen
  \bibfield  {author} {\bibinfo {author} {\bibfnamefont {Tiago~P.}\
  \bibnamefont {Peixoto}},\ }\bibfield  {title} {\enquote {\bibinfo {title}
  {Latent {Poisson} models for networks with heterogeneous density},}\ }\href
  {http://arxiv.org/abs/2002.07803} {\bibfield  {journal} {\bibinfo  {journal}
  {arXiv:2002.07803 [physics, stat]}\ } (\bibinfo {year} {2020})},\ \bibinfo
  {note} {arXiv: 2002.07803}\BibitemShut {NoStop}%
\bibitem [{\citenamefont {Metropolis}\ \emph {et~al.}(1953)\citenamefont
  {Metropolis}, \citenamefont {Rosenbluth}, \citenamefont {Rosenbluth},
  \citenamefont {Teller},\ and\ \citenamefont
  {Teller}}]{metropolis_equation_1953}%
  \BibitemOpen
  \bibfield  {author} {\bibinfo {author} {\bibfnamefont {Nicholas}\
  \bibnamefont {Metropolis}}, \bibinfo {author} {\bibfnamefont {Arianna~W.}\
  \bibnamefont {Rosenbluth}}, \bibinfo {author} {\bibfnamefont {Marshall~N.}\
  \bibnamefont {Rosenbluth}}, \bibinfo {author} {\bibfnamefont {Augusta~H.}\
  \bibnamefont {Teller}}, \ and\ \bibinfo {author} {\bibfnamefont {Edward}\
  \bibnamefont {Teller}},\ }\bibfield  {title} {\enquote {\bibinfo {title}
  {Equation of {State} {Calculations} by {Fast} {Computing} {Machines}},}\
  }\href {\doibase 10.1063/1.1699114} {\bibfield  {journal} {\bibinfo
  {journal} {The Journal of Chemical Physics}\ }\textbf {\bibinfo {volume}
  {21}},\ \bibinfo {pages} {1087} (\bibinfo {year} {1953})}\BibitemShut
  {NoStop}%
\bibitem [{\citenamefont {Hastings}(1970)}]{hastings_monte_1970}%
  \BibitemOpen
  \bibfield  {author} {\bibinfo {author} {\bibfnamefont {W.~K.}\ \bibnamefont
  {Hastings}},\ }\bibfield  {title} {\enquote {\bibinfo {title} {Monte {Carlo}
  sampling methods using {Markov} chains and their applications},}\ }\href
  {\doibase 10.1093/biomet/57.1.97} {\bibfield  {journal} {\bibinfo  {journal}
  {Biometrika}\ }\textbf {\bibinfo {volume} {57}},\ \bibinfo {pages} {97 --109}
  (\bibinfo {year} {1970})}\BibitemShut {NoStop}%
\bibitem [{\citenamefont {Kunegis}(2013)}]{kunegis_konect:_2013}%
  \BibitemOpen
  \bibfield  {author} {\bibinfo {author} {\bibfnamefont {Jérôme}\
  \bibnamefont {Kunegis}},\ }\bibfield  {title} {\enquote {\bibinfo {title}
  {{KONECT}: {The} {Koblenz} {Network} {Collection}},}\ }in\ \href {\doibase
  10.1145/2487788.2488173} {\emph {\bibinfo {booktitle} {Proceedings of the
  {22Nd} {International} {Conference} on {World} {Wide} {Web}}}},\ \bibinfo
  {series and number} {{WWW} '13 {Companion}}\ (\bibinfo  {publisher} {ACM},\
  \bibinfo {address} {New York, NY, USA},\ \bibinfo {year} {2013})\ pp.\
  \bibinfo {pages} {1343--1350}\BibitemShut {NoStop}%
\bibitem [{\citenamefont
  {Peixoto}(2014{\natexlab{c}})}]{peixoto_graph-tool_2014}%
  \BibitemOpen
  \bibfield  {author} {\bibinfo {author} {\bibfnamefont {Tiago~P.}\
  \bibnamefont {Peixoto}},\ }\bibfield  {title} {\enquote {\bibinfo {title}
  {The \texttt{graph-tool} python library},}\ }\href {\doibase
  10.6084/m9.figshare.1164194} {\bibfield  {journal} {\bibinfo  {journal}
  {figshare}\ } (\bibinfo {year} {2014}{\natexlab{c}}),\
  10.6084/m9.figshare.1164194},\ \bibinfo {note} {available at
  \url{https://graph-tool.skewed.de}.}\BibitemShut {Stop}%
\bibitem [{\citenamefont {Girvan}\ and\ \citenamefont
  {Newman}(2002)}]{girvan_community_2002}%
  \BibitemOpen
  \bibfield  {author} {\bibinfo {author} {\bibfnamefont {M.}~\bibnamefont
  {Girvan}}\ and\ \bibinfo {author} {\bibfnamefont {M.~E.~J.}\ \bibnamefont
  {Newman}},\ }\bibfield  {title} {\enquote {\bibinfo {title} {Community
  structure in social and biological networks},}\ }\href {\doibase
  10.1073/pnas.122653799} {\bibfield  {journal} {\bibinfo  {journal}
  {Proceedings of the National Academy of Sciences}\ }\textbf {\bibinfo
  {volume} {99}},\ \bibinfo {pages} {7821 --7826} (\bibinfo {year}
  {2002})}\BibitemShut {NoStop}%
\bibitem [{\citenamefont {Jain}\ and\ \citenamefont
  {Neal}(2004)}]{jain_split-merge_2004}%
  \BibitemOpen
  \bibfield  {author} {\bibinfo {author} {\bibfnamefont {Sonia}\ \bibnamefont
  {Jain}}\ and\ \bibinfo {author} {\bibfnamefont {Radford~M.}\ \bibnamefont
  {Neal}},\ }\bibfield  {title} {\enquote {\bibinfo {title} {A {Split}-{Merge}
  {Markov} chain {Monte} {Carlo} {Procedure} for the {Dirichlet} {Process}
  {Mixture} {Model}},}\ }\href {\doibase 10.1198/1061860043001} {\bibfield
  {journal} {\bibinfo  {journal} {Journal of Computational and Graphical
  Statistics}\ }\textbf {\bibinfo {volume} {13}},\ \bibinfo {pages} {158--182}
  (\bibinfo {year} {2004})}\BibitemShut {NoStop}%
\bibitem [{\citenamefont {Jain}\ and\ \citenamefont
  {Neal}(2007)}]{jain_splitting_2007}%
  \BibitemOpen
  \bibfield  {author} {\bibinfo {author} {\bibfnamefont {Sonia}\ \bibnamefont
  {Jain}}\ and\ \bibinfo {author} {\bibfnamefont {Radford~M.}\ \bibnamefont
  {Neal}},\ }\bibfield  {title} {{\selectlanguage {english}\enquote {\bibinfo
  {title} {Splitting and merging components of a nonconjugate {Dirichlet}
  process mixture model},}\ }}\href {\doibase 10.1214/07-BA219} {\bibfield
  {journal} {\bibinfo  {journal} {Bayesian Analysis}\ }\textbf {\bibinfo
  {volume} {2}},\ \bibinfo {pages} {445--472} (\bibinfo {year}
  {2007})}\BibitemShut {NoStop}%
\bibitem [{\citenamefont {Wang}\ and\ \citenamefont
  {Russell}(2015)}]{wang_smart-dumb/dumb-smart_2015}%
  \BibitemOpen
  \bibfield  {author} {\bibinfo {author} {\bibfnamefont {Wei}\ \bibnamefont
  {Wang}}\ and\ \bibinfo {author} {\bibfnamefont {Stuart}\ \bibnamefont
  {Russell}},\ }\bibfield  {title} {\enquote {\bibinfo {title} {A
  smart-dumb/dumb-smart algorithm for efficient split-merge {MCMC}},}\ \
  }(\bibinfo  {publisher} {AUAI Press},\ \bibinfo {year} {2015})\ pp.\ \bibinfo
  {pages} {902--911}\BibitemShut {NoStop}%
\bibitem [{\citenamefont {Moody}(2001)}]{moody_peer_2001}%
  \BibitemOpen
  \bibfield  {author} {\bibinfo {author} {\bibfnamefont {James}\ \bibnamefont
  {Moody}},\ }\bibfield  {title} {\enquote {\bibinfo {title} {Peer influence
  groups: identifying dense clusters in large networks},}\ }\href@noop {}
  {\bibfield  {journal} {\bibinfo  {journal} {Social Networks}\ }\textbf
  {\bibinfo {volume} {23}},\ \bibinfo {pages} {261--283} (\bibinfo {year}
  {2001})}\BibitemShut {NoStop}%
\bibitem [{\citenamefont {Knuth}(1993)}]{knuth_stanford_1993}%
  \BibitemOpen
  \bibfield  {author} {\bibinfo {author} {\bibfnamefont {Donald~E.}\
  \bibnamefont {Knuth}},\ }\href@noop {} {\emph {\bibinfo {title} {The
  {Stanford} {GraphBase}: {A} {Platform} for {Combinatorial} {Computing}}}},\
  \bibinfo {edition} {1st}\ ed.\ (\bibinfo  {publisher} {Addison-Wesley
  Professional},\ \bibinfo {address} {New York, N.Y. : Reading, Mass},\
  \bibinfo {year} {1993})\BibitemShut {NoStop}%
\bibitem [{\citenamefont {White}\ \emph {et~al.}(1986)\citenamefont {White},
  \citenamefont {Southgate}, \citenamefont {Thomson},\ and\ \citenamefont
  {Brenner}}]{white_structure_1986}%
  \BibitemOpen
  \bibfield  {author} {\bibinfo {author} {\bibfnamefont {J.~G.}\ \bibnamefont
  {White}}, \bibinfo {author} {\bibfnamefont {E.}~\bibnamefont {Southgate}},
  \bibinfo {author} {\bibfnamefont {J.~N.}\ \bibnamefont {Thomson}}, \ and\
  \bibinfo {author} {\bibfnamefont {S.}~\bibnamefont {Brenner}},\ }\bibfield
  {title} {{\selectlanguage {english}\enquote {\bibinfo {title} {The structure
  of the nervous system of the nematode {Caenorhabditis} elegans},}\
  }}\href@noop {} {\bibfield  {journal} {\bibinfo  {journal} {Philosophical
  Transactions of the Royal Society of London. Series B, Biological Sciences}\
  }\textbf {\bibinfo {volume} {314}},\ \bibinfo {pages} {1--340} (\bibinfo
  {year} {1986})}\BibitemShut {NoStop}%
\bibitem [{\citenamefont {Newman}(2006)}]{newman_finding_2006}%
  \BibitemOpen
  \bibfield  {author} {\bibinfo {author} {\bibfnamefont {M.~E.~J.}\
  \bibnamefont {Newman}},\ }\bibfield  {title} {\enquote {\bibinfo {title}
  {Finding community structure in networks using the eigenvectors of
  matrices},}\ }\href {\doibase 10.1103/PhysRevE.74.036104} {\bibfield
  {journal} {\bibinfo  {journal} {Physical Review E}\ }\textbf {\bibinfo
  {volume} {74}},\ \bibinfo {pages} {036104} (\bibinfo {year}
  {2006})}\BibitemShut {NoStop}%
\bibitem [{\citenamefont {Larremore}\ \emph {et~al.}(2013)\citenamefont
  {Larremore}, \citenamefont {Clauset},\ and\ \citenamefont
  {Buckee}}]{larremore_network_2013}%
  \BibitemOpen
  \bibfield  {author} {\bibinfo {author} {\bibfnamefont {Daniel~B.}\
  \bibnamefont {Larremore}}, \bibinfo {author} {\bibfnamefont {Aaron}\
  \bibnamefont {Clauset}}, \ and\ \bibinfo {author} {\bibfnamefont
  {Caroline~O.}\ \bibnamefont {Buckee}},\ }\bibfield  {title} {\enquote
  {\bibinfo {title} {A {Network} {Approach} to {Analyzing} {Highly}
  {Recombinant} {Malaria} {Parasite} {Genes}},}\ }\href {\doibase
  10.1371/journal.pcbi.1003268} {\bibfield  {journal} {\bibinfo  {journal}
  {PLOS Comput Biol}\ }\textbf {\bibinfo {volume} {9}},\ \bibinfo {pages}
  {e1003268} (\bibinfo {year} {2013})}\BibitemShut {NoStop}%
\bibitem [{\citenamefont {Adamic}\ and\ \citenamefont
  {Glance}(2005)}]{adamic_political_2005}%
  \BibitemOpen
  \bibfield  {author} {\bibinfo {author} {\bibfnamefont {Lada~A.}\ \bibnamefont
  {Adamic}}\ and\ \bibinfo {author} {\bibfnamefont {Natalie}\ \bibnamefont
  {Glance}},\ }\bibfield  {title} {\enquote {\bibinfo {title} {The political
  blogosphere and the 2004 {U}.{S}. election: divided they blog},}\ }in\ \href
  {\doibase 10.1145/1134271.1134277} {\emph {\bibinfo {booktitle} {Proceedings
  of the 3rd international workshop on {Link} discovery}}},\ \bibinfo {series
  and number} {{LinkKDD} '05}\ (\bibinfo  {publisher} {ACM},\ \bibinfo
  {address} {New York, NY, USA},\ \bibinfo {year} {2005})\ pp.\ \bibinfo
  {pages} {36--43}\BibitemShut {NoStop}%
\end{thebibliography}%

\end{document}